\documentclass[preprint, 3p, sort&compress]{elsarticle}

\usepackage{amssymb}
\usepackage{amsthm}
\usepackage[mathlines]{lineno}
\usepackage{graphicx}
\usepackage{units}
\usepackage{url}
\usepackage{amsmath}
\usepackage{amsfonts}
\usepackage{bm}
\usepackage{textcomp}
\usepackage{subfigure}
\usepackage{multicol}
\usepackage{multirow}
\usepackage{verbatim}
\usepackage{rotating}
\usepackage[colorlinks,linkcolor=black,citecolor=red]{hyperref}
\usepackage{float}
\usepackage{epsfig}
\usepackage{dcolumn}
\usepackage{bm}
\usepackage{color}
\usepackage{pstricks}
\usepackage{pst-node}
\usepackage{times}
\usepackage{indentfirst}
\usepackage[english]{babel}
\usepackage{epstopdf}
\addto{\captionsenglish}{%

}
\journal{Physics Letters B}


\usepackage[font=footnotesize]{caption}

\usepackage{overpic}

\newcommand{\ev}{\,\mbox{eV}}
\newcommand{\mev}{\,\mbox{MeV}}

\newcommand{\mevcc}{\,\mbox{MeV}/c^2}

\newcommand{\gev}{\,\mbox{GeV}}

\newcommand{\gevcc}{\,\mbox{GeV}/c^2}

\newcommand{\pio}{\pi^0}

\newcommand{\ee}{e^+e^-}

\biboptions{numbers,sort&compress}

\begin{document}

%\pagewiselinenumbers
%\linenumbers
\begin{frontmatter}

\title{\boldmath Observation of a resonant structure in $e^{+}e^{-} \to \omega\eta$ and another in $e^{+}e^{-} \to \omega\pi^{0}$ at center-of-mass energies between 2.00 and 3.08 GeV}
\author{
%\begin{small}
%\begin{center}
M.~Ablikim$^{1}$, M.~N.~Achasov$^{10,d}$, P.~Adlarson$^{64}$, S.~Ahmed$^{15}$, M.~Albrecht$^{4}$, A.~Amoroso$^{63A,63C}$, Q.~An$^{60,48}$, ~Anita$^{21}$, Y.~Bai$^{47}$, O.~Bakina$^{29}$
% ...
\\
\vspace{0.2cm}
(BESIII Collaboration)\\
\vspace{0.2cm} %{\it
$^{1}$ Institute of High Energy Physics, Beijing 100049, People's Republic of China\\
% ...
$^{4}$ Bochum Ruhr-University, D-44780 Bochum, Germany\\
% ...
$^{10}$ G.I. Budker Institute of Nuclear Physics SB RAS (BINP), Novosibirsk 630090, Russia\\
% ...
$^{15}$ Helmholtz Institute Mainz, Johann-Joachim-Becher-Weg 45, D-55099 Mainz, Germany\\
% ...
$^{21}$ Indian Institute of Technology Madras, Chennai 600036, India\\
% ...
$^{29}$ Joint Institute for Nuclear Research, 141980 Dubna, Moscow region, Russia\\
% ...
$^{47}$ Southeast University, Nanjing 211100, People's Republic of China\\
$^{48}$ State Key Laboratory of Particle Detection and Electronics, Beijing 100049, Hefei 230026, People's Republic of China\\
% ...
$^{60}$ University of Science and Technology of China, Hefei 230026, People's Republic of China\\
% ...
$^{63}$ (A)University of Turin, I-10125, Turin, Italy; (B)University of Eastern Piedmont, I-15121, Alessandria, Italy; (C)INFN, I-10125, Turin, Italy\\
$^{64}$ Uppsala University, Box 516, SE-75120 Uppsala, Sweden\\
% ...
$^{d}$ Also at the Novosibirsk State University, Novosibirsk, 630090, Russia\\
% ...
%}
%\end{center}
\vspace{0.4cm}
%\end{small}
}
%\noaffiliation{}

\date{\today}

\begin{abstract}
Born cross sections for the processes $e^+e^- \to \omega\eta$ and
$e^+e^- \to \omega\pi^{0}$ have been determined for center-of-mass
energies between 2.00 and 3.08$\gev$ with the BESIII detector at the
BEPCII collider. The results obtained in this work are consistent with
previous measurements but with improved precision. Two resonant structures
are observed.  In the $\ee \to \omega\eta$ cross sections, a resonance
with a mass of $(2179 \pm 21 \pm 3)\mevcc$ and a width of $(89 \pm 28 \pm 5)\mev$ is observed with a significance of
6.1$\sigma$. Its properties are consistent with the $\phi(2170)$.  In
the $\ee \to\omega\pio$ cross sections, a resonance denoted $Y(2040)$
is observed with a significance of more than 10$\sigma$. Its mass and
width are determined to be $(2034 \pm 13 \pm 9)\mevcc$ and $(234 \pm 30 \pm 25)\mev$, respectively, where the first uncertainties are
statistical and the second ones are systematic.
\end{abstract}

\end{frontmatter}

%\maketitle

\begin{multicols}{2}

\section{Introduction}
In low-energy $e^+e^-$ collision experiments, the vector mesons
$\rho$, $\omega$, and $\phi$ and their low lying excited states can be
produced abundantly.  The Particle Data Group (PDG)~\cite{PDG} has
tabulated experimental results for these states. However, some of
the higher lying excitations are not fully identified yet. It is
especially in the region around 2~GeV where further experimental
insight is needed to resolve the situation involving resonances such
as the $\rho(2000)$, $\rho(2150)$ and $\phi(2170)$ states.

Considerable efforts have been made theoretically to understand the
nature of the $\phi(2170)$ resonance, and several interpretations have
been proposed, such as an $s\bar s g$ hybrid~\cite{ssg_phi2170, ssg2_phi2170}, 
an $s\bar s$ meson~\cite{ssbar_phi2170, ssbar2_phi2170, ssbar3_phi2170, ssbar4_phi2170}, 
an $s\bar s s\bar s$ tetraquark
state~\cite{sssbarsbar1_phi2170,sssbarsbar2_phi2170,sssbarsbar3_phi2170, sssbarsbar4_phi2170, sssbarsbar5_phi2170, sssbarsbar6_phi2170}, a $\Lambda\bar
\Lambda$ bound state~\cite{Lambda1_phi2170,Lambda2_phi2170,
  Lambda3_phi2170}, as well as $\phi K\bar K$~\cite{phiKK_phi2170} and
$\phi f_0(980)$~\cite{phif02_phi2170} resonances.  These models differ
in their predictions of the branching fractions of the $\phi(2170)$ to
decay channels such as $\phi\eta$ or $K^{(*)}\bar{K}^{(*)}$ as certain
decay modes can either be suppressed or favored depending on its
nature~\cite{ssg_phi2170, ssbar_phi2170, quarkonia, LLbarbound,
  phietatheory}.  It is therefore of great importance to measure the
branching fractions for a variety of different decay channels in order
to help in discriminating between different models.

The $\phi(2170)$ state was first observed by the BaBar experiment in
the initial state radiation (ISR) process
$\ee\to\gamma_\mathrm{ISR}\phi f_0(980)$~\cite{BABAR_phi2170} and
later confirmed by the BESII and BESIII experiments in
$J/\psi\to\eta\phi f_0(980)$~\cite{BESII_phi2170, BESIII_phi2170} as
well as by both the BaBar and Belle experiments in the aforementioned
ISR process~\cite{BABAR2_phi2170, Belle_phi2170}.  The observed masses
and widths of the $\phi(2170)$ range from
$(2079\pm13^{+79}_{-28})\mevcc$~\cite{BABAR2_phi2170} to
$(2200\pm6\pm5)\mevcc$~\cite{BESIII_phi2170} and
$(58\pm16\pm20)\mev$~\cite{BABAR_phi2170} to
$(192\pm23^{+25}_{-61})\mev$~\cite{Belle_phi2170}, respectively.

Several studies of the properties of the $\phi(2170)$ resonance have
recently been made by the BESIII experiment.  A partial-wave analysis
was performed for the $e^+e^- \to K^+K^-\pio\pio$
process~\cite{BESKKpipi}, in which indications for sizable partial
widths of the $\phi(2170)$ resonance to the $K^+(1460)K^-$,
$K_1^+(1270)K^-$ and $K_1^+(1400)K^-$ channels (here,
charge-conjugation is implied) were found.  Attempts were also made to
study channels with simpler topologies, including
% 
% % Nils: they quote something here that has not yet been published but still needs spokesperson approval - hopefully this resolves before this document is published, otherwise it might be best to completely remove this reference?
%
$e^{+}e^{-}\to K^{+}K^{-}$, where a resonance with mass
$(2239.2\pm7.1\pm11.3 )\mevcc$ and width $(139.8\pm12.3\pm20.6)\mev$
was found~\cite{BESKK, BESKKTheory}, and
$e^{+}e^{-}\to\phi\eta^\prime$~\cite{BESphietap}, where a resonance
with mass $(2177.5~\pm~5.1~\pm~18.6)\mevcc$ and width
$(149.0~\pm~15.6~\pm~8.9)\mev$ was found, In $e^{+}e^{-}\to\phi
K^{+}K^{-}$, a sharp enhancement is observed in the Born cross
section at $\sqrt{s}=2.2324$ GeV, which is close to the mass of the $
\phi(2170)$ resonance~\cite{BESphiKK}, however its width seems to be
incompatible with that of the $\phi(2170)$.

A comparison of decay channels without hidden or open strangeness such as $e^+e^- \to \omega \eta$ to those observed thus far can provide additional information about the properties of the $\phi(2170)$ resonance.  
In addition, this process can also be used to study excited $\omega$ resonances appearing as $\omega^*\to\omega\eta$~\cite{omegafamily}, 
which is expected to be one of the dominant decay channels for excited $\omega$ mesons and a benchmark process to study their properties.
%{\color{red} isnot $\rho\pi$ the dominant decay mode?}

In contrast to the $\ee \to \omega \eta$ process, the reaction
$\ee\to\omega\pi^0$ allows the study of the isovector vector mesons
and their excited states. Generally, the excited $\rho$ states around
$2~\gevcc$ are not well understood.  Although there are two results on the
so-called $\rho(2000)$~\cite{rho1, rho2}, its existence is not
well-established.
Furthermore, several experiments have claimed the observation of the
$\rho(2150)$ state with mass and width lying in the range of
$1.990$ to $2.254\gevcc$ and $70$ to $389\mev$,
respectively~\cite{babar2020, rho3, rho4, rho5, rho6}.

In an approach based on the quark-pair-creation model,
the $\rho(2150)$ state is identified as a candidate for the $4^3S_1$
state~\cite{Rhoresonance2,Rhoresonance}.
The Born cross section of $e^+e^-\to\omega\pi^0$ in the
energy region below 2~GeV has been measured by several
experiments~\cite{SND2000,SND2003,SND2016,ND,CMD1,CMD2,BABAR,DM2},
while the data above 2~GeV is rather scarce.
Thus, more measurements of $e^+e^-\to\omega\pi^0$ above 2~GeV are of
high interest to study the properties of excited $\rho$
states.

In this letter, we present Born cross section measurements of the processes $e^+e^- \to \omega\eta$ and $e^+e^- \to \omega\pi^{0}$ with subsequent  $\omega \to \pi^+\pi^-\pi^0$, $\pi^0 \to \gamma\gamma$ and $\eta \to \gamma \gamma$ decays.

\section{Detector and data sample}
%%%%%%%%%%%%MC simulation%%%%%%%%%%%%%%%%%%%%%%%%%%%%%%%%%%%%%%%%%%%%%%%%%%%%%%%%%%%%%%%%%%%
The BESIII detector is a magnetic spectrometer~\cite{Ablikim:2009aa} located at the Beijing Electron
Position Collider (BEPCII)~\cite{Yu:IPAC2016-TUYA01}. The cylindrical core of the BESIII detector consists of a helium-based
 multilayer drift chamber (MDC), a plastic scintillator time-of-flight system (TOF), and a CsI(Tl) electromagnetic calorimeter (EMC),
which are all enclosed in a superconducting solenoidal magnet providing a 1.0~T 
magnetic field. The solenoid is supported by an octagonal flux-return yoke with resistive plate counter muon identifier modules interleaved with steel.
The acceptance of charged particles and photons is 93\% over $4\pi$ solid angle. The charged-particle momentum resolution at $1~{\rm GeV}/c$ is $0.5\%$, and the $dE/dx$ resolution is $6\%$ for the electrons from Bhabha scattering. The EMC measures photon energies with a resolution of $2.5\%$ ($5\%$) at $1$~GeV in the barrel (end cap) region. The time resolution of the TOF barrel part is 68~ps, while that of the end cap part is 110~ps.

The data samples used in this letter have been collected with the BESIII detector at 22 center-of-mass (c.m.) energies from 2.000 to 3.080 $\gev$, corresponding to a total integrated luminosity of 651 pb$^{-1}$.

The {\sc geant4} based~\cite{geant4} simulation software {\sc boost}~\cite{boost} is used to produce Monte Carlo (MC) simulation samples.
Events are generated using the \textsc{ConExc} generator~\cite{ConExc} with ISR and vacuum polarization (VP) taken into account.
Inclusive hadron production of the type $e^+e^- \to$ hadrons is simulated to estimate possible background processes and to optimize event selection criteria.
Exclusive MC samples are generated to determine the detection efficiencies of the signal processes. 
Since the beam energy spread of BEPCII is less than 1~MeV at $\sqrt{s}< 3~{\rm GeV}$, it is much smaller than the experimental resolution of the BESIII detector and can thus be ignored in the simulation.

\section{Event selection and determination of the Born cross section}
\subsection{Analysis of $e^+e^- \to \omega \eta$ }
\label{selomegaeta}
%%%%%%%%%%%%Event selection for e^+e^- \to omega eta %%%%%%%%%%%%%%%%%%%%%%%%%%%%%%%%%%%%%%%%%
For $e^+e^- \to \omega \eta$ (with subsequent $\omega \to
\pi^+\pi^-\pi^0$, $\pi^0 \to \gamma\gamma$ and $\eta \to \gamma
\gamma$ decays), candidate events are required to have at least two
reconstructed charged tracks and at least four reconstructed photons.
Each charged track is required to be located within the MDC
acceptance, $|\cos\theta| < 0.93$, where $\theta$ is the polar angle
of the charged track, and to originate from a cylinder around the
interaction point of 1 cm radius and extending $\pm10$~cm along the detector axis. 
Information from TOF and $dE/dx$
measurements is combined to form particle identification (PID)
likelihoods for the $\pi$, $K$, and $p$ hypotheses.  Each track is
assigned a particle type corresponding to the hypothesis with the
highest PID likelihood.  Exactly two oppositely charged pions are
required in each event.  Photon candidates are reconstructed using
clusters of energy deposited in the EMC crystals.  The energy is
required to be larger than $25\mev$ in the barrel region
($|\cos\theta| < 0.80$) and larger than $50\mev$ in the end cap region
($0.86< |\cos\theta| < 0.92$).  The energy deposited in nearby TOF
counters is included to improve the reconstruction efficiency and
energy resolution.  The difference of the EMC time from the event
start time is required to be within [0,700] ns to suppress electronic
noise and showers unrelated to the event.

To improve the momentum and energy resolution and to suppress
background events, a four-constraint (4C) kinematic fit imposing
four-momentum conservation is performed under the hypothesis $e^+e^-
\to \pi^{+}\pi^{-}4\gamma$. For the goodness of the kinematic fit,
$\chi^{2}_{\rm 4C} < 70$ is required.  For events with more than four
photon candidates, the combination with the smallest $\chi^{2}_{\rm
  4C}$ is retained.  In addition, a kinematic fit for the alternative
hypothesis $e^+e^- \to \pi^{+}\pi^{-}5\gamma$ is performed and only
those events that satisfy $\chi_{\rm 4C}^{2}(\pi^{+}~\pi^{-}4\gamma) <
\chi_{\rm 4C}^{2}(\pi^{+}\pi^{-}5\gamma)$ are retained in order to
suppress backgrounds from $\ee \to \omega\pi^{0}\pi^{0}$ events.  Two
photon pairs corresponding to the best $\pi^{0}\eta$, $\pi^{0}\pi^{0}$
and $\eta\eta$ candidates are selected separately by choosing the
combination with the smallest value of
$\chi^{2}_{\alpha\beta}=(M(\gamma_1\gamma_2)-m_{\alpha})^{2}/\sigma^{2}_{12}
+ (M(\gamma_3\gamma_4)-m_{\beta}^{2})/\sigma^{2}_{34}$, where $\alpha$
and $\beta$ represent either $\pi^0$ or $\eta$, and the mass
resolution $\sigma_{12(34)}$ in the invariant mass region of the
$\pi^0$ or $\eta$ meson is obtained from MC simulations.  Only
combinations with $\chi^{2}_{\pi^{0}\eta} < \chi^{2}_{\pi^{0}\pi^{0}}$
and $\chi^{2}_{\pi^{0}\eta} < \chi^{2}_{\eta\eta}$ are retained.  The
$\pi^{0}$ and $\eta$ candidates are selected by requiring
$|M(\gamma_1\gamma_2)-m_{\pi^{0}}|< 0.02\gevcc$ and
$|M(\gamma_3\gamma_4)-m_{\eta}|< 0.03\gevcc$, corresponding to about
$3\sigma$ intervals around the respective nominal masses of $\pi^{0}$
and $\eta$, $m_{\pi^{0}}$ and $m_{\eta}$~\cite{PDG}.  Events with
$|E_{\gamma_{3}} - E_{\gamma_{4}}|/p_{\eta} > 0.9$, where $p_{\eta}$
is the momentum of the $\eta$ meson in the laboratory system, are
rejected to suppress background events from the $\ee \to
\omega\gamma_{\rm ISR}$ and $\ee \to \omega\pi^{0}\pi^{0}$ processes.

\end{multicols}

\begin{figure}[htbp]
\begin{center}
\begin{overpic}[width=0.46\textwidth]{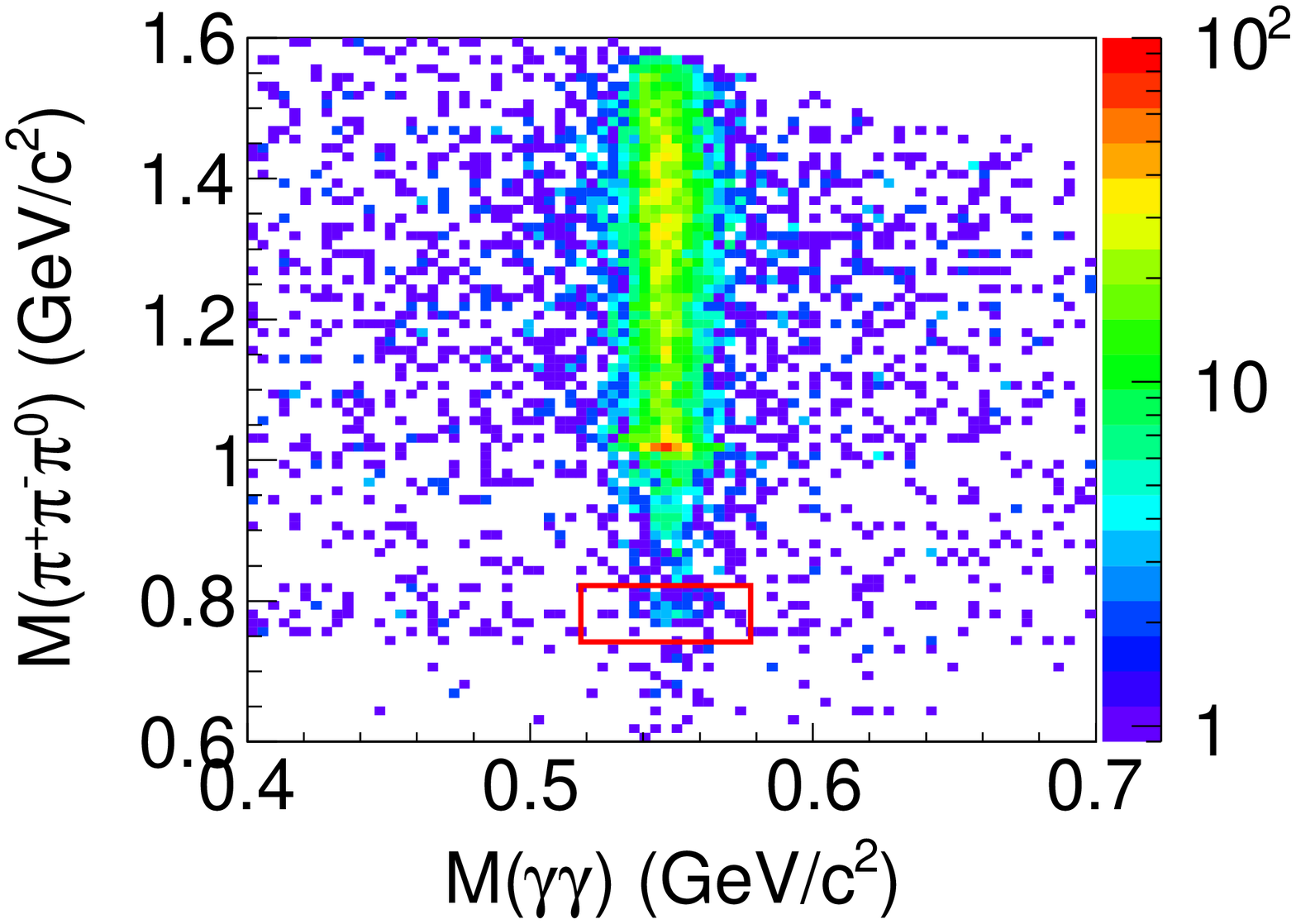}
\put(75,64){(a)}
\end{overpic}
\begin{overpic}[width=0.46\textwidth]{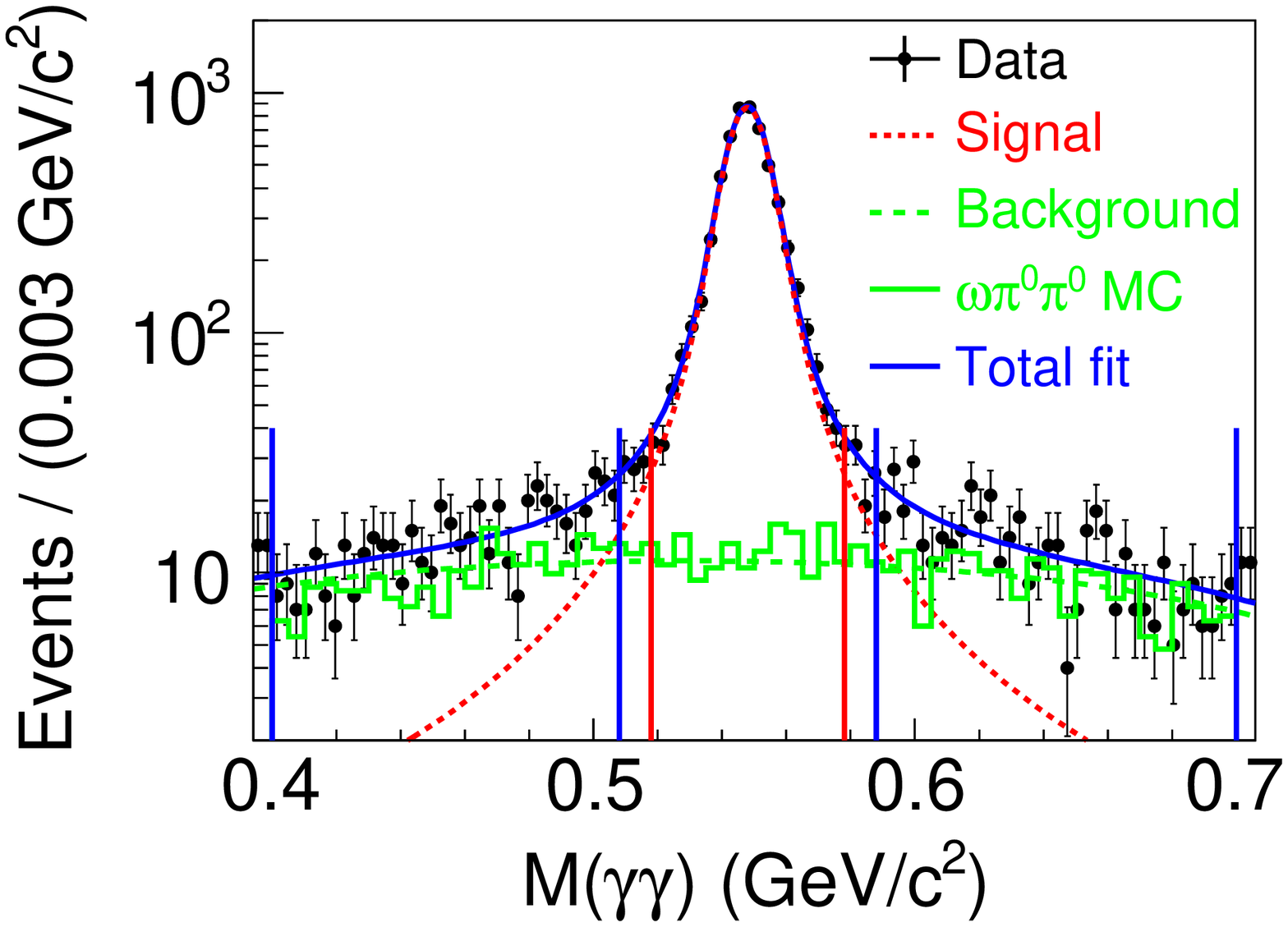}
\put(28,64){(b)}
\end{overpic}

\begin{overpic}[width=0.46\textwidth]{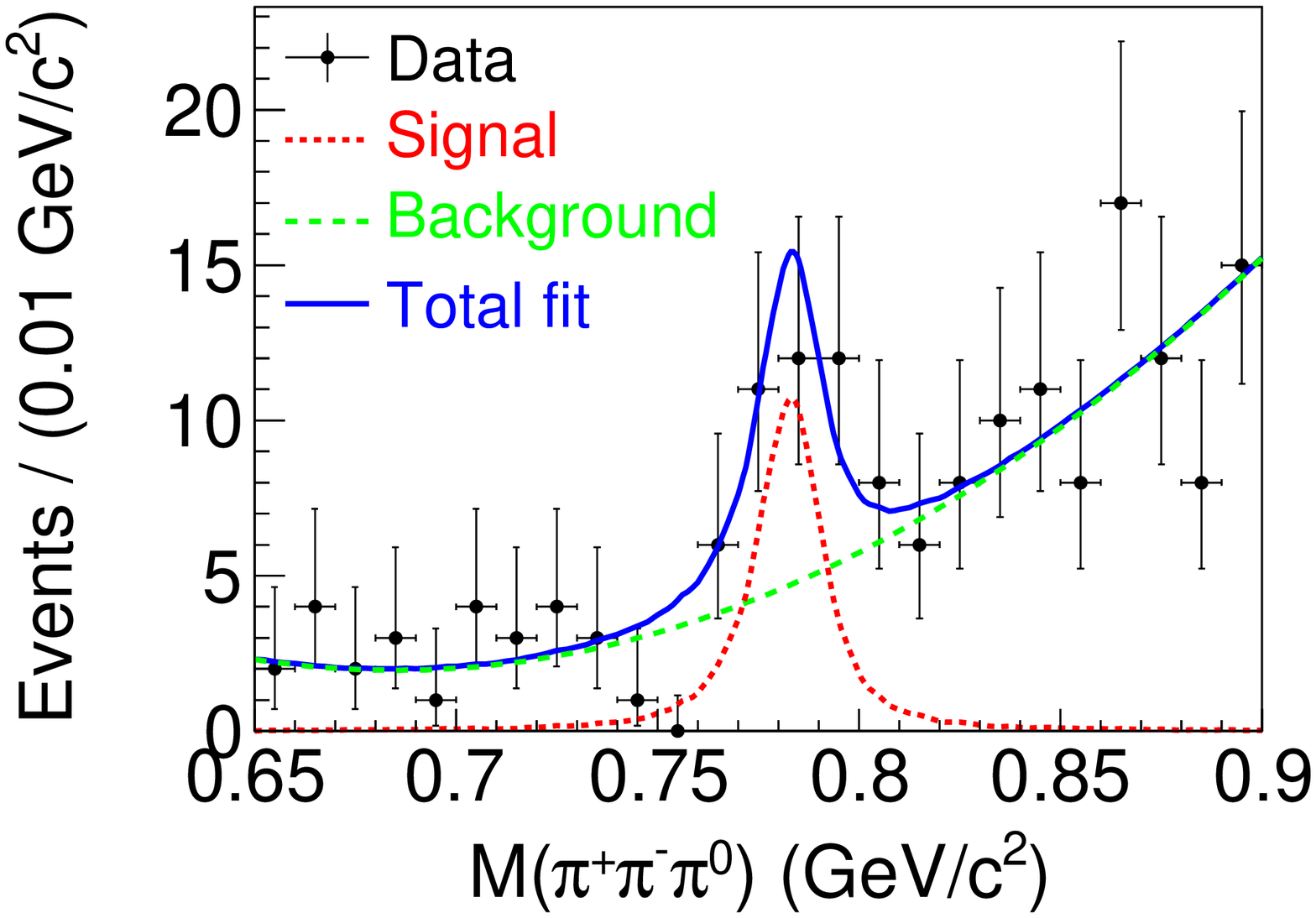}
\put(80,64){(c)}
\end{overpic}
\begin{overpic}[width=0.46\textwidth]{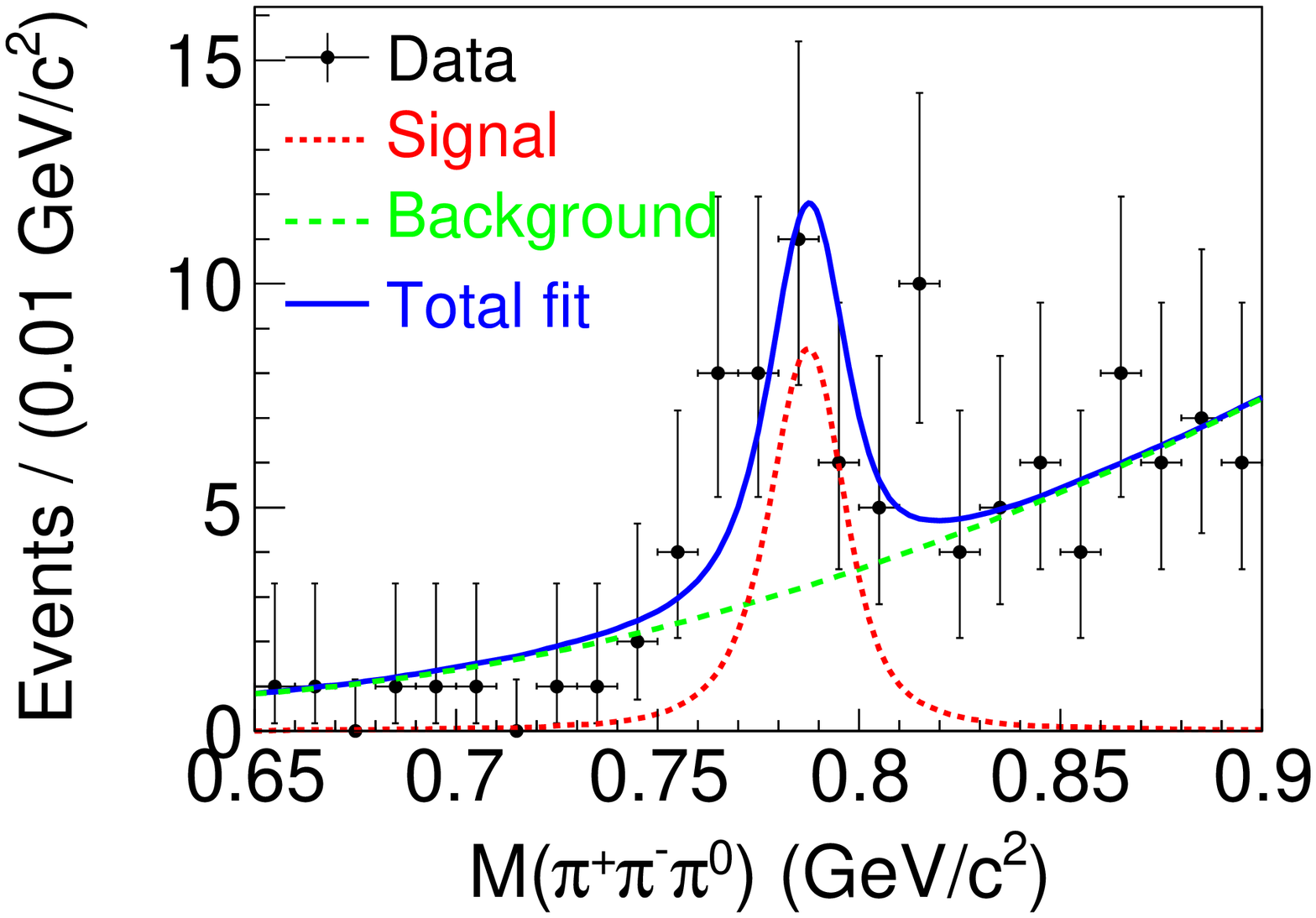}
\put(80,64){(d)}
\end{overpic}
\end{center}
\vspace{-0.5cm}
\caption{(Color online) Invariant mass distributions for data taken at
  $\sqrt{s}=2.125$ GeV. (a) Distribution of the $\pi^+\pi^-\pi^0$
  invariant mass versus the two-photon invariant mass. The
  area marked in red corresponds to the signal region.  (b) Fit to the
  $M(\gamma\gamma)$ distribution, where the (black) dots with error
  bars are data, the (blue) solid curve is the total fit result, the
  (green) dashed curve indicates background described by a second
  order Chebychev polynomial, the (red) dotted curve is the
  $\eta\to\gamma\gamma$ signal shape described by a Voigt function and
  the (green) histogram is the $\ee\to\omega\pi^0\pi^0$ MC sample
  scaled to the integral of the background function in the fit.  The
  vertical lines indicate the signal (red) and sideband regions
  (blue).  (c) and (d) represent the $M(\pi^{+}\pi^{-}\pi^{0})$
  invariant mass distributions in the $\eta$ signal and sideband
  region, respectively.  The dots with error bars are data, the solid
  curves are the total fit results, the dashed curves indicate the
  background described by a second order Chebychev polynomial and the
  dotted curves are the $\omega$ signal shapes determined from MC
  simulations convolved with a Gaussian accounting for a potential
  difference in resolution between data and MC simulation.}
\label{OmegaFit}
\end{figure}

\begin{multicols}{2}

The distribution of the $\pi^{+}\pi^{-}\pi^0$ invariant mass versus the two-photon invariant mass of the selected events at
$\sqrt{s}=2.125$ GeV is shown as an example in Fig.~\ref{OmegaFit}(a), where an $\omega$ signal around the nominal $\omega$ meson mass
is visible.  %The accumulation of events around
%$M(\pi^{+}\pi^{-}\pi^0)=1.02\gevcc$ is due to the
%$\phi\to\pi^+\pi^-\pi^0$ decay.
%above not visible in figure. fah
Potential background reactions to the $e^+e^- \to \omega \eta$ process
are studied using both inclusive $e^+e^- \to$ hadrons and exclusive MC
samples.  Simulated events are subject to the same selection procedure
as that applied to the experimental data.  According to MC
simulations, the dominant background stems from $e^+e^- \to
\pi^+\pi^-\pi^0\eta$, which contains the same final state particles as
the signal reaction.  The $\ee\to\omega\pi^0\pi^0$ and
$\ee\to\omega\gamma_{\rm ISR}$ processes form a peaking background
contribution in the $\pi^{+}\pi^{-}\pi^0$ invariant mass distribution.
The total peaking background from $\ee\to\omega\gamma_{\rm ISR}$ is
estimated by MC simulations normalized to the experimental luminosity
and is found to be negligible.  The peaking background from
$\ee\to\omega\pi^0\pi^0$ is inferred from the $\eta$
sidebands, which are defined as $0.400 < M(\gamma_3\gamma_4) <
0.508 \gevcc$ and $0.588 < M(\gamma_3\gamma_4) < 0.700 \gevcc$ as
shown in Fig.~\ref{OmegaFit}(b).

To determine the signal yield of the $e^+e^- \to \omega\eta$ process, a simultaneous unbinned maximum likelihood fit is performed to the $M(\pi^+\pi^-\pi^0)$ spectra in both the $\eta$ signal and background regions at each energy, where the shapes of signal and background are shared. 
%{\color{red}
Figure~\ref{OmegaFit}~(c) and (d) show the fit results in signal and sideband regions at 2.125 GeV. 
%}
The signal is modeled with the peak shape obtained from MC simulation convolved with a Gaussian function allowing for a potential resolution difference between data and MC simulation.
The background is described with a second-order Chebychev polynomial.
In the fit, peaking background is automatically subtracted by constructing the number of $\omega$ events in the $\eta$ signal region as $N_{\rm obs}=N_{\rm sig}+f_{\rm scale}\cdot N_{\rm bkg}$, where $N_{\rm sig}$ is the number of $\omega\eta$ signal events, $N_{\rm bkg}$ is the number of $\omega$ events in the $\eta$ sideband region, and 
%$f_{\rm scale}$ is a scale factor determined by the ratio of background events in the $\eta$ signal and sideband regions.
%{\color{red} 
%$f_{\rm scale}$ is a scale factor determined by the ratio of non-$\eta$ events in the $\eta$ signal and sideband regions shown Fig.~\ref{OmegaFit}(b).
%}
$f_{\rm scale}$ is the normalization factor 
$f_{\rm scale} = N^{{\eta}_{\rm sig}}/N^{{\eta}_{\rm sideband}}$
where $N^{{\eta}_{\rm sig}}$($N^{{\eta}_{\rm sideband}}$) is the number of background events falling into the signal (sideband)
region as shown in Fig.~\ref{OmegaFit}(b).

\begin{figure}[H]%[htbp]
\begin{center}
\begin{overpic}[width=0.492\textwidth]{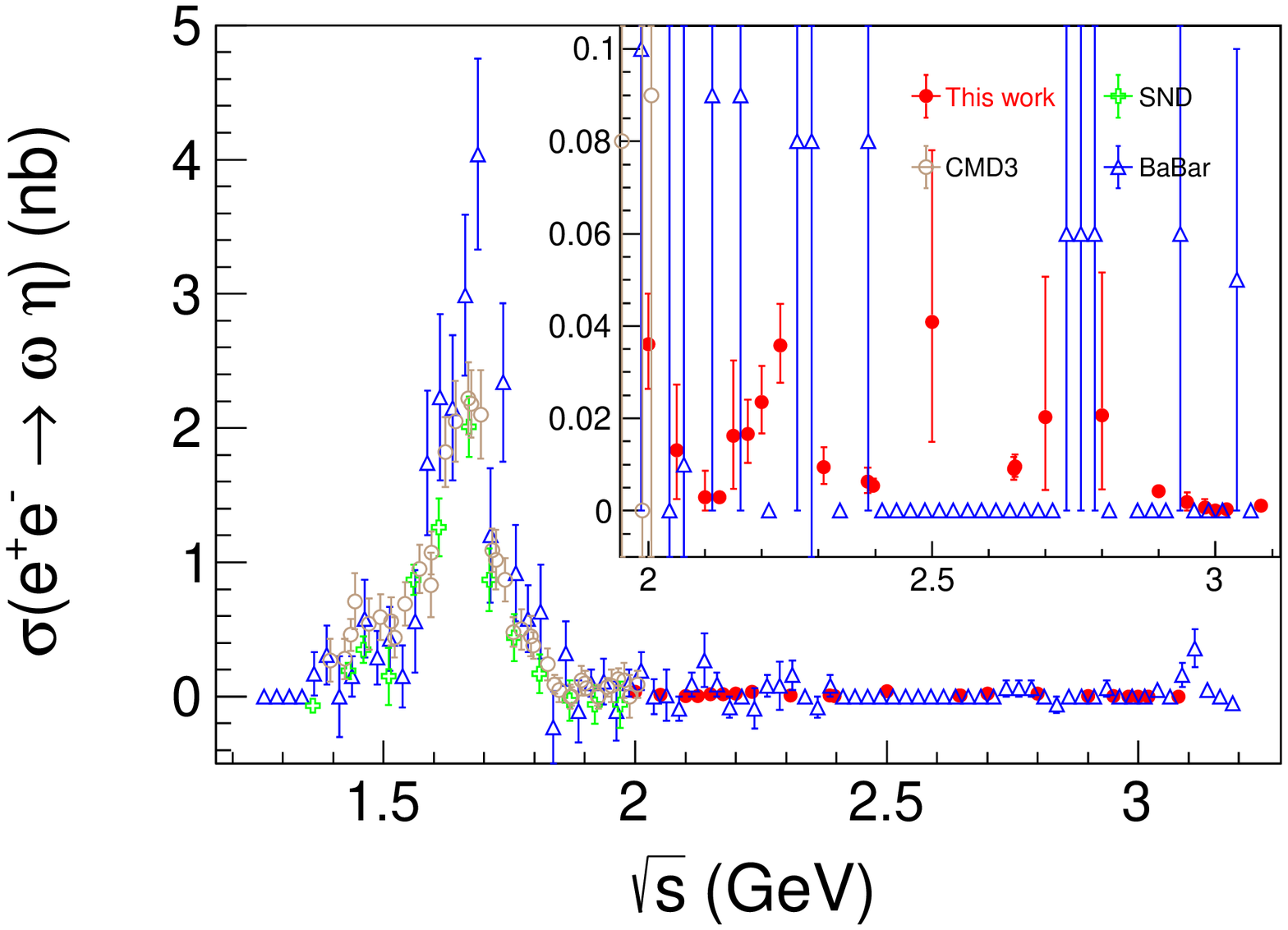}
\put(22,63){(a)}
\end{overpic}
\begin{overpic}[width=0.492\textwidth]{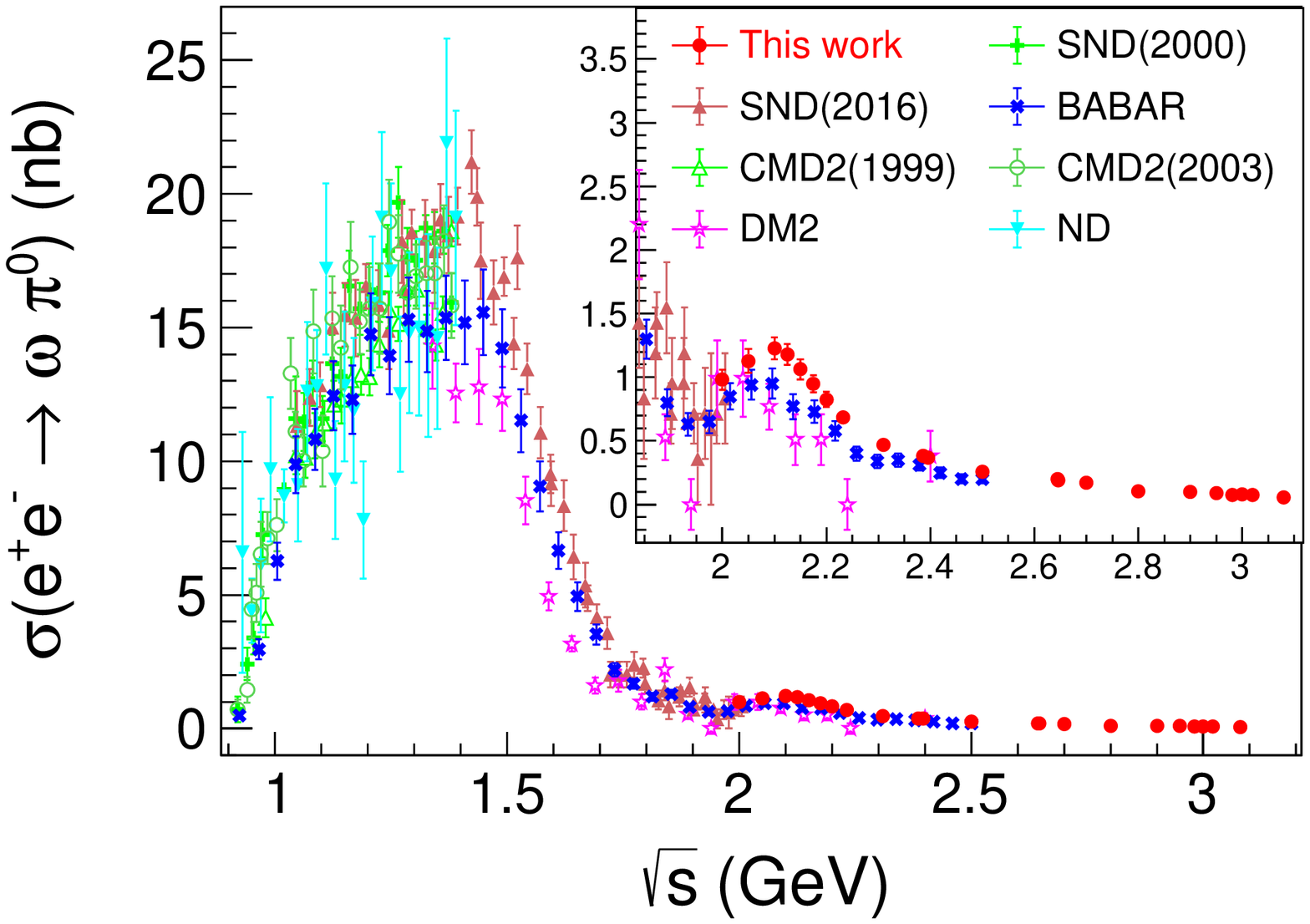}
\put(22,63){(b)}
\end{overpic}
\vspace{-0.5cm}
\caption{(Color online) Dressed cross sections for the processes (a) $\ee\to\omega\eta$ and (b) $\ee\to\omega\pi^0$. In comparison to the data presented in this work (red dots), in (a) the data from the CMD3~\cite{CMD2017omegaeta} (brown open circles), SND~\cite{SND2016omegaeta} (green open crosses) and BaBar~\cite{BaBarOmegaEta} (blue open triangles) experiments are shown. In (b), our data is compared to the results of the CMD2~\cite{CMD1,CMD2} (green open upward triangles and green open circles), SND~\cite{SND2000,SND2016} (green filled crosses and brown filled triangles), BaBar~\cite{BABAR} (blue filled X crosses), DM2~\cite{DM2} (magenta open stars) and ND~\cite{ND} (cyan filled downward triangles) experiments. }\label{fig:comparison}
\end{center}
\end{figure}

The Born cross section of the $e^+e^- \to \omega\eta$ process is calculated according to

\begin{equation}
\begin{split}
&\sigma = \frac{N_{\rm sig}}{\mathcal{L}\cdot\varepsilon\cdot(1+\delta)\cdot\mathcal{B}},\\
\end{split}
\label{CXOE}
\end{equation}
%where $N_{\rm sig}$ is the number of signal events. %determined by $N_{\rm sig} = N_{\rm obs} - f_{\rm scale}\cdot N_{\rm bkg}$. 
where $\mathcal{L}$ is the integrated luminosity of the individual dataset, ($1 +\delta$) is the radiative correction factor accounting for both ISR and VP, and $\epsilon$ is the product of geometrical acceptance and selection efficiency obtained from MC simulation.
The total branching fraction $\mathcal{B}$ is the product of the branching fractions for the decays contained in the full decay chain $\mathcal{B}=\mathcal{B}(\omega \to \pi^+\pi^-\pi^0)\cdot\mathcal{B}(\pi^0 \to \gamma\gamma)\cdot\mathcal{B}(\eta \to \gamma\gamma)=34.7\%$.
The Born cross sections as well as upper limits at the 90\% confidence level are given for all 22 energy points together with all values used in the calculation in Table~\ref{crossOmegaEta}. 
VP factors are also listed for the convenience of calculating dressed cross sections. 
The results are consistent with previous measurements~\cite{SND2016omegaeta, CMD2017omegaeta, BaBarOmegaEta} but with improved precision. 
A comparison to the previous results is shown in Fig.~\ref{fig:comparison}(a).

Various sources of systematic uncertainties concerning the measurement
of the Born cross sections are investigated, including integrated luminosity,
branching fractions, ISR and VP correction factors, event selection
criteria, the fit procedure of the signal, and the contributions from
peaking background processes.

\end{multicols}

%%%%%%%%%%%%%%%%%%%%%%%
\begin{table}[H] %[htbp]
\setlength{\tabcolsep}{10pt}
  \centering
  \footnotesize
  \caption{The Born cross sections of the $e^{+}e^{-} \to \omega \eta$ process. In addition, upper limits are given at 90\% confidence level. All symbols defined are the same as those in Eq.~(\ref{CXOE}). 
In the column of Born cross section $\sigma$, the first uncertainty is statistical, and the second one is systematic. 
Sig. is the significance of the observed signal. VP lists the vacuum polarization factor.
}
%\resizebox{0.65\textwidth}{!}{\begin{minipage}{\textwidth}
    \label{crossOmegaEta}
\begin{tabular}{c c c c c c c c c c}
  \hline
  \hline
$\sqrt{s}$ (GeV) &$N_{\rm sig}$ & $N^{\rm up}_{\rm sig}$  &$\mathcal{L}$ (${\rm pb}^{-1}$)  &$\varepsilon\cdot(1+\delta)$ &$\sigma$ (pb) & $\sigma^{\rm up}$ (pb) & Sig. ($\sigma$) & VP\\
\hline
2.0000 & 19.3$^{+5.9}_{-5.2}$  &  $<$27.3 & 10.1 & 0.158 & 34.7$^{+10.6}_{-9.3}\pm2.9$ & $<$49.3 &  4.5 & 1.037  \\ 
2.0500 & 2.3$^{+2.6}_{-1.9}$  &  $<$7.0 & 3.34 & 0.161 & 12.6$^{+13.7}_{-10.1}\pm1.0$ & $<$37.5 &  1.4 & 1.038  \\ 
2.1000 & 1.9$^{+3.8}_{-1.9}$  &  $<$8.2 & 12.2 & 0.162 & 2.8$^{+5.6}_{-2.8}\pm0.2$ & $<$11.9 &  0.5 & 1.039  \\ 
2.1250 & 17.2$^{+8.2}_{-7.5}$  &  $<$26.0 & 108 & 0.163 & 2.8$^{+1.3}_{-1.2}\pm0.3$ & $<$4.3 &  2.2 & 1.039  \\ 
2.1500 & 2.3$^{+2.3}_{-1.6}$  &  $<$6.0 & 2.84 & 0.151 & 15.6$^{+15.7}_{-11.0}\pm0.7$ & $<$40.3 &  1.1 & 1.040  \\ 
2.1750 & 9.2$^{+4.1}_{-3.4}$  &  $<$14.9 & 10.6 & 0.156 & 16.0$^{+7.1}_{-6.0}\pm1.0$ & $<$25.9 &  3.0 & 1.040  \\ 
2.2000 & 16.5$^{+5.5}_{-4.8}$  &  $<$25.0 & 13.7 & 0.153 & 22.7$^{+7.5}_{-6.5}\pm1.7$ & $<$34.3 &  4.3 & 1.040  \\ 
2.2324 & 22.9$^{+5.8}_{-5.1}$  &  $<$30.9 & 11.9 & 0.161 & 34.4$^{+8.7}_{-7.7}\pm2.2$ & $<$46.4 &  $>$5 & 1.041  \\ 
2.3094 & 11.9$^{+5.3}_{-4.6}$  &  $<$22.6 & 21.1 & 0.178 & 9.1$^{+4.1}_{-3.5}\pm0.7$ & $<$17.3 &  3.7 & 1.041  \\ 
2.3864 & 8.2$^{+3.9}_{-3.3}$  &  $<$14.5 & 22.5 & 0.173 & 6.1$^{+2.9}_{-2.4}\pm0.4$ & $<$10.7 &  2.6 & 1.041  \\ 
2.3960 & 20.6$^{+6.3}_{-5.6}$  &  $<$29.6 & 66.9 & 0.172 & 5.2$^{+1.6}_{-1.4}\pm0.4$ & $<$7.4 &  3.5 & 1.041  \\ 
2.5000 & 2.6$^{+2.4}_{-1.7}$  &  $<$6.3 & 1.10 & 0.175 & 39.3$^{+35.7}_{-25.0}\pm3.5$ & $<$94.2 &  1.6 & 1.041  \\ 
2.6444 & 17.7$^{+5.2}_{-4.5}$  &  $<$23.3 & 33.7 & 0.174 & 8.7$^{+2.6}_{-2.2}\pm0.5$ & $<$11.4 &  $>$5 & 1.039  \\ 
2.6464 & 18.8$^{+5.1}_{-4.4}$  &  $<$26.0 &  34.0 & 0.173 & 9.2$^{+2.5}_{-2.2}\pm0.6$ & $<$12.7 &  $>$5 & 1.039  \\ 
2.7000 & 1.2$^{+1.9}_{-1.0}$  &  $<$2.2 & 1.03 & 0.177 & 19.6$^{+29.3}_{-15.2}\pm0.9$ & $<$34.7 &  1.1 & 1.039  \\ 
2.8000 & 1.2$^{+1.9}_{-1.0}$  &  $<$2.2 & 1.01 & 0.177 & 20.0$^{+29.8}_{-15.5}\pm0.9$ & $<$35.4 &  1.1 & 1.037  \\ 
2.9000 & 27.0$^{+6.0}_{-5.3}$  &  $<$30.3 & 105 & 0.182 & 4.1$^{+0.9}_{-0.8}\pm0.3$ & $<$4.6 &  $>$5 & 1.033  \\ 
2.9500 & 1.8$^{+2.1}_{-1.8}$  &  $<$5.0 & 15.9 & 0.184 & 1.8$^{+2.1}_{-1.7}\pm0.1$ & $<$4.9 &  0.7 & 1.029  \\ 
2.9810 & 0.7$^{+1.8}_{-0.7}$  &  $<$4.4 & 16.1 & 0.187 & 0.7$^{+1.8}_{-0.7}\pm0.1$ & $<$4.2 &  0.2 & 1.025  \\ 
3.0000 & 0.0$^{+0.5}_{-0.0}$  &  $<$2.2 & 15.9 & 0.186 & 0.0$^{+0.5}_{-0.0}\pm0.0$ & $<$2.1 &  0.0 & 1.021  \\ 
3.0200 & 0.3$^{+1.4}_{-0.3}$  &  $<$2.2 & 17.3 & 0.184 & 0.3$^{+1.3}_{-0.3}\pm0.0$ & $<$2.0 &  0.3 & 1.014  \\ 
3.0800 & 9.2$^{+4.5}_{-3.8}$  &  $<$15.8 & 126 & 0.172 & 1.2$^{+0.6}_{-0.5}\pm0.1$ & $<$2.1 &  2.8 & 0.915  \\
\hline
\hline
    \end{tabular}
%\end{minipage}}
\end{table}

\begin{multicols}{2}

The integrated luminosity at each energy point is measured using large
angle Bhabha events with an uncertainty of 1\% following the method in
Ref.~\cite{Luminosity}.  The uncertainties associated with the
branching fractions of intermediate states are taken from the
PDG~\cite{PDG}.  
%The uncertainty in the ISR and VP correction is estimated as described in Ref.~\cite{ISRcorrection}.  
%
The uncertainty of the ISR and VP correction factors is obtained from the accuracy of radiation function, which is about 0.5\%~\cite{ConExc}, and has an additional contribution from the cross section lineshape, 
which is estimated by varying the model parameters of the fit to the cross sections. All parameters are randomly varied within their uncertainties and the resulting parametrization of the lineshape is used to recalculate $(1+\delta)$, $\epsilon$ and the corresponding cross sections. This procedure is repeated 1000 times and the standard deviation of the resulting cross sections is taken as a systematic uncertainty.
%
%In the simulation of the signal process, the generator calculates $(1+\delta)$ from the lineshape, which is extracted from experimental results by fitting the cross sections with a specific model. 
%There are uncertainties in the parameters of the model, and its influence to the cross section measurement can be estimated by varying the parameters considering their uncertainties and correlations. 
%
Differences between the data and MC simulation for the tracking efficiency and PID
of charged pions are investigated using the high-purity control sample
of $e^{+}e^{-} \to K^{+}K^{-}\pi^{+}\pi^{-}$~\cite{Tracking, BESKK}.  The photon
detection efficiency is studied with a sample of $e^{+}e^{-} \to
K^{+}K^{-}\pi^{+}\pi^{-}\pi^{0}$ 
%{\color{red} 
with similar method for tracking uncertainty~\cite{Tracking}.
%} 
The result shows that the
difference in detection efficiency between data and MC simulation is
1\% per photon. 
The uncertainties associated with the kinematic fit
are studied with the track helix parameter correction method, as
described in Ref.~\cite{4C}.

\end{multicols}

\begin{table}[H]
%\begin{sidewaystable}
\setlength{\tabcolsep}{6.6pt}
  \centering
  \scriptsize
  \caption{
  Summary of relative systematic uncertainties (in \%) associated with the luminosity ($\mathcal{L}$),  the tracking efficiency (Track), the photon detection efficiency (Photon), PID, Branching fraction (Br), $\chi^2$ requirement, 4C kinematic fit (4C),  $|E_{\gamma_{3}} - E_{\gamma_{4}}|/p_{\eta} <0.9$ (Angle), background shape (Bkg), signal shape (Sig), fit range (Range), $\eta$ and $\pi^0$ mass windows (m($\eta$) and m($\pi^0$)), peaking background (Peak), the initial state radiation and the vacuum polarization correction factor ($1+\delta$) in the measurement of the Born cross section of the  $e^{+}e^{-} \to \omega \eta$ process.}
 \label{UncertaintyOE}
  \begin{tabular}{ccccc ccccc ccccc cc}
  \hline
  \hline
Energies &$\mathcal{L}$ &Track &Photon &PID &Br &$\chi^{2}$ &4C &Angle  &Bkg &Sig  &Range   & m($\eta$)    &m($\pi^{0}$)    &Peak & $1+\delta$  &Total    \\
\hline
 2.0000  & 1.0  & 2.0  & 4.0  & 2.0  & 0.9  & 0.5  & 0.6  & 1.1  & 0.1  & 2.5  & 0.2  & 0.4  & 3.0  & 4.8  & 0.5  & 8.1  \\ 
 2.0500  & 1.0  & 2.0  & 4.0  & 2.0  & 0.9  & 0.5  & 0.6  & 1.1  & 0.1  & 2.5  & 0.2  & 0.4  & 3.0  & 4.8  & 0.6  & 8.2  \\ 
 2.1000  & 1.0  & 2.0  & 4.0  & 2.0  & 0.9  & 0.5  & 0.6  & 1.1  & 0.1  & 2.5  & 0.2  & 0.4  & 3.0  & 4.8  & 2.7  & 8.6  \\ 
 2.1250  & 1.0  & 2.0  & 4.0  & 2.0  & 0.9  & 0.5  & 0.5  & 1.1  & 1.9  & 2.5  & 2.2  & 0.4  & 3.0  & 9.6  & 1.1  & 12  \\ 
 2.1500  & 1.0  & 2.0  & 4.0  & 2.0  & 0.9  & 0.5  & 0.6  & 1.1  & 0.1  & 2.5  & 0.2  & 0.4  & 3.0  & 4.8  & 1.5  & 8.3  \\ 
 2.1750  & 1.0  & 2.0  & 4.0  & 2.0  & 0.9  & 0.5  & 0.7  & 1.1  & 0.5  & 2.5  & 0.5  & 0.4  & 3.0  & 3.7  & 1.2  & 7.7  \\ 
 2.2000  & 1.0  & 2.0  & 4.0  & 2.0  & 0.9  & 0.5  & 0.5  & 1.1  & 0.3  & 1.6  & 0.2  & 0.4  & 3.0  & 2.6  & 1.8  & 7.0  \\ 
 2.2324  & 1.0  & 2.0  & 4.0  & 2.0  & 0.9  & 0.5  & 0.7  & 1.1  & 0.2  & 0.8  & 0.7  & 0.4  & 3.0  & 1.6  & 1.5  & 6.5  \\ 
 2.3094  & 1.0  & 2.0  & 4.0  & 2.0  & 0.9  & 0.5  & 0.7  & 1.1  & 0.0  & 2.0  & 0.1  & 0.4  & 3.0  & 2.1  & 0.7  & 6.8  \\ 
 2.3864  & 1.0  & 2.0  & 4.0  & 2.0  & 0.9  & 0.5  & 0.6  & 1.1  & 1.0  & 1.6  & 1.2  & 0.4  & 3.0  & 2.3  & 0.5  & 6.9  \\ 
 2.3960  & 1.0  & 2.0  & 4.0  & 2.0  & 0.9  & 0.5  & 0.5  & 1.1  & 0.0  & 2.2  & 0.2  & 0.4  & 3.0  & 2.6  & 0.5  & 7.0  \\ 
 2.5000  & 1.0  & 2.0  & 4.0  & 2.0  & 0.9  & 0.5  & 0.6  & 1.1  & 0.1  & 2.5  & 0.2  & 0.4  & 3.0  & 4.8  & 0.5  & 8.1  \\ 
 2.6444  & 1.0  & 2.0  & 4.0  & 2.0  & 0.9  & 0.5  & 0.6  & 1.1  & 0.2  & 1.4  & 0.6  & 0.4  & 3.0  & 1.6  & 0.5  & 6.5  \\ 
 2.6464  & 1.0  & 2.0  & 4.0  & 2.0  & 0.9  & 0.5  & 0.6  & 1.1  & 0.1  & 0.8  & 0.8  & 0.4  & 3.0  & 1.7  & 0.5  & 6.4  \\ 
 2.7000  & 1.0  & 2.0  & 4.0  & 2.0  & 0.9  & 0.5  & 0.6  & 1.1  & 0.1  & 2.5  & 0.2  & 0.4  & 3.0  & 4.8  & 0.5  & 8.1  \\ 
 2.8000  & 1.0  & 2.0  & 4.0  & 2.0  & 0.9  & 0.5  & 0.6  & 1.1  & 0.1  & 2.5  & 0.2  & 0.4  & 3.0  & 4.8  & 0.5  & 8.1  \\ 
 2.9000  & 1.0  & 2.0  & 4.0  & 2.0  & 0.9  & 0.5  & 0.5  & 1.1  & 0.2  & 0.4  & 0.8  & 0.4  & 3.0  & 1.7  & 0.5  & 6.4  \\ 
 2.9500  & 1.0  & 2.0  & 4.0  & 2.0  & 0.9  & 0.5  & 0.6  & 1.1  & 0.1  & 2.5  & 0.2  & 0.4  & 3.0  & 4.8  & 0.5  & 8.1  \\ 
 2.9810  & 1.0  & 2.0  & 4.0  & 2.0  & 0.9  & 0.5  & 0.6  & 1.1  & 0.1  & 2.5  & 0.2  & 0.4  & 3.0  & 4.8  & 0.5  & 8.1  \\ 
 3.0000  & 1.0  & 2.0  & 4.0  & 2.0  & 0.9  & 0.5  & 0.6  & 1.1  & 0.1  & 2.5  & 0.2  & 0.4  & 3.0  & 4.8  & 0.5  & 8.1  \\ 
 3.0200  & 1.0  & 2.0  & 4.0  & 2.0  & 0.9  & 0.5  & 0.6  & 1.1  & 0.1  & 2.5  & 0.2  & 0.4  & 3.0  & 4.8  & 0.5  & 8.1  \\ 
 3.0800  & 1.0  & 2.0  & 4.0  & 2.0  & 0.9  & 0.5  & 0.7  & 1.1  & 0.4  & 1.6  & 0.9  & 0.4  & 3.0  & 2.7  & 0.5  & 6.9  \\
 \hline
  \hline
    \end{tabular}
%\end{sidewaystable}
\end{table}
%%%%%%%%%%%%%%%%%%%%%%%

\begin{multicols}{2}

\noindent
 Due to the limited statistics in the
data samples, a control sample of the $J/\psi \to \omega\eta$ decay is
used to estimate the uncertainties arising from the selection
conditions $\chi_{\rm 4C}^{2}(\pi^{+}~\pi^{-}4\gamma) < \chi_{\rm
  4C}^{2}(\pi^{+}~\pi^{-}5\gamma)$, $\chi_{\pi^{0}\eta}^{2} <
\chi_{\pi^{0}\pi^{0}}^{2}$, $\chi_{\pi^{0}\eta}^{2} <
\chi_{\eta\eta}^{2}$, $|M(\gamma_1\gamma_2)-m_{\pi^{0}}|< 0.02\gevcc$,
$|M(\gamma_3\gamma_4)-m_{\eta}|< 0.03\gevcc$ and $|E_{\gamma_{3}} -
E_{\gamma_{4}}|/p_{\eta} <0.9$.
 For this, the single-requirement
efficiency is studied, removing one of the selection conditions at a
time and studying the change in the number of observed events. In case
a significant difference is found between the data control sample and
a MC simulation of the $J/\psi \to \omega\eta$ decay, this difference
is taken as the systematic uncertainty.

Due to large statistical fluctuations in the data, toy MC samples are used to estimate the systematic uncertainties stemming from the description of the signal and background shape as well as from the fit range when determining $N_{\text{obs}}$.
A total of 500 sets of toy MC samples are generated according to the final fit result shown in Fig.~\ref{OmegaFit}(c) with the same statistics as in data.
For each toy MC sample, 
the following procedure is performed:
the $\omega$ signal shape is changed to a Breit-Wigner function convolved with a Gaussian, 
the background shape is varied from a second to a third order Chebychev polynomial and the fit range is varied by $\pm10 \mevcc$. 
The mean value of the differences of the signal yield between the nominal and the alternative fits is taken 
as the systematic uncertainty. 
%This procedure is repeated for the peaking background reactions contained in the $\eta$ sideband.
%{\color{red}
The uncertainty of peaking background is related to the uncertainty of $N_{\text{bkg}}$ and $f_{\text{scale}}$.
We estimate uncertainty of $N_{\text{bkg}}$ with the same method for $N_{\text{obs}}$, and that of $f_{\text{scale}}$ by considering the fit uncertainty of the non-$\eta$ background at 2.125 GeV.
%The peaking background is from events corresponding to non-$\eta$ which is estimated from $\eta$ sideband, 
%Since in Fig.~\ref{OmegaFit} (d) non-eta event 
%}

The total systematic uncertainty for the Born cross section measurement is determined to be 12\% for the $e^+e^- \to \omega\eta$ process at $\sqrt{s}=2.125$ GeV. The uncertainties at the other c.m. energies are determined accordingly and are summarized in Table~\ref{UncertaintyOE}.

\subsection{Analysis of $e^+e^- \to \omega \pi^0$}
%%%%%%%%%%%%Event selection for e^+e^- \to omega pi0%%%%%%%%%%%%%%%%%%%%%%%%%%%%%%%%%%%%%%%%%
The event selection criteria for the $e^+e^- \to \omega \pi^0$ process are mostly the same as described in Sec.~\ref{selomegaeta}. 
The $\pi^{0}\pi^{0}$ candidate pairs are selected by minimizing $\chi^{2}_{\pi^0\pi^0}=(M(\gamma_1\gamma_2)-m_{\pi^0})^{2}/\sigma^{2}_{12} + (M(\gamma_3\gamma_4)-m_{\pi^0})^{2}/\sigma^{2}_{34}$. These $\pi^{0}$ candidates are required to be in a mass window of $(m_{\pi^{0}}- 0.02\gevcc, m_{\pi^{0}}+ 0.02\gevcc)$. 
Since there are two $\pi^{0}$ candidates, the $\pi^{+}\pi^{-}\pi^0$ combination whose invariant mass is closest to $m_{\omega}$ is retained as the $\omega$ candidate, where the $\pi^0$ is denoted as $\pi^0_\omega$ to distinguish it from the bachelor pion $\pi^0_\mathrm{bach}$.

\begin{figure}[htbp]
\begin{center}
%\subfigure{\includegraphics[width=0.32\textwidth]{graph/Omega_eta/2125Omega_dis.eps}}
\begin{overpic}[width=0.46\textwidth]{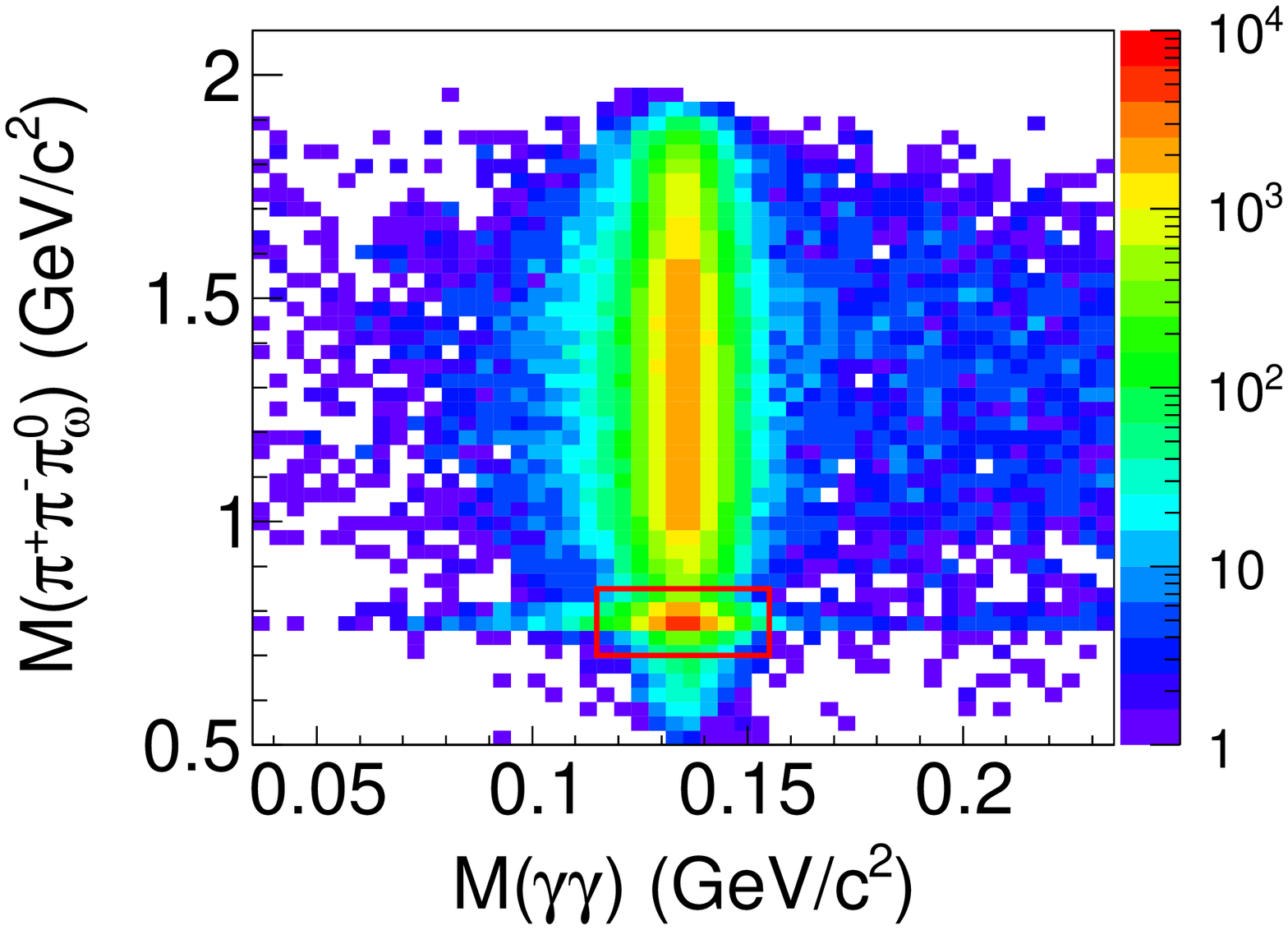}
\put(28,60){(a)}
\end{overpic}
\begin{overpic}[width=0.46\textwidth]{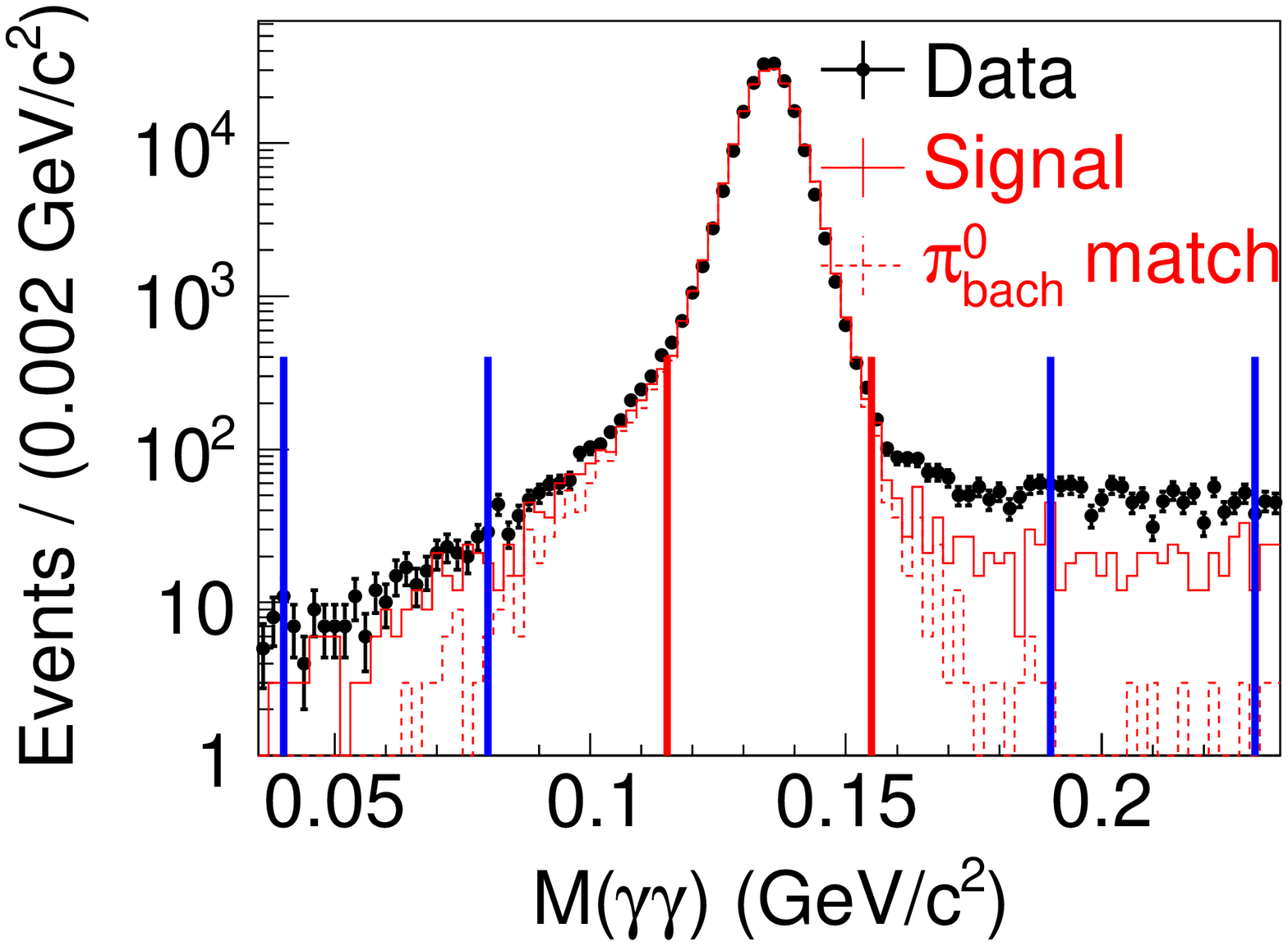}
\put(28,60){(b)}
\end{overpic}

\begin{overpic}[width=0.46\textwidth]{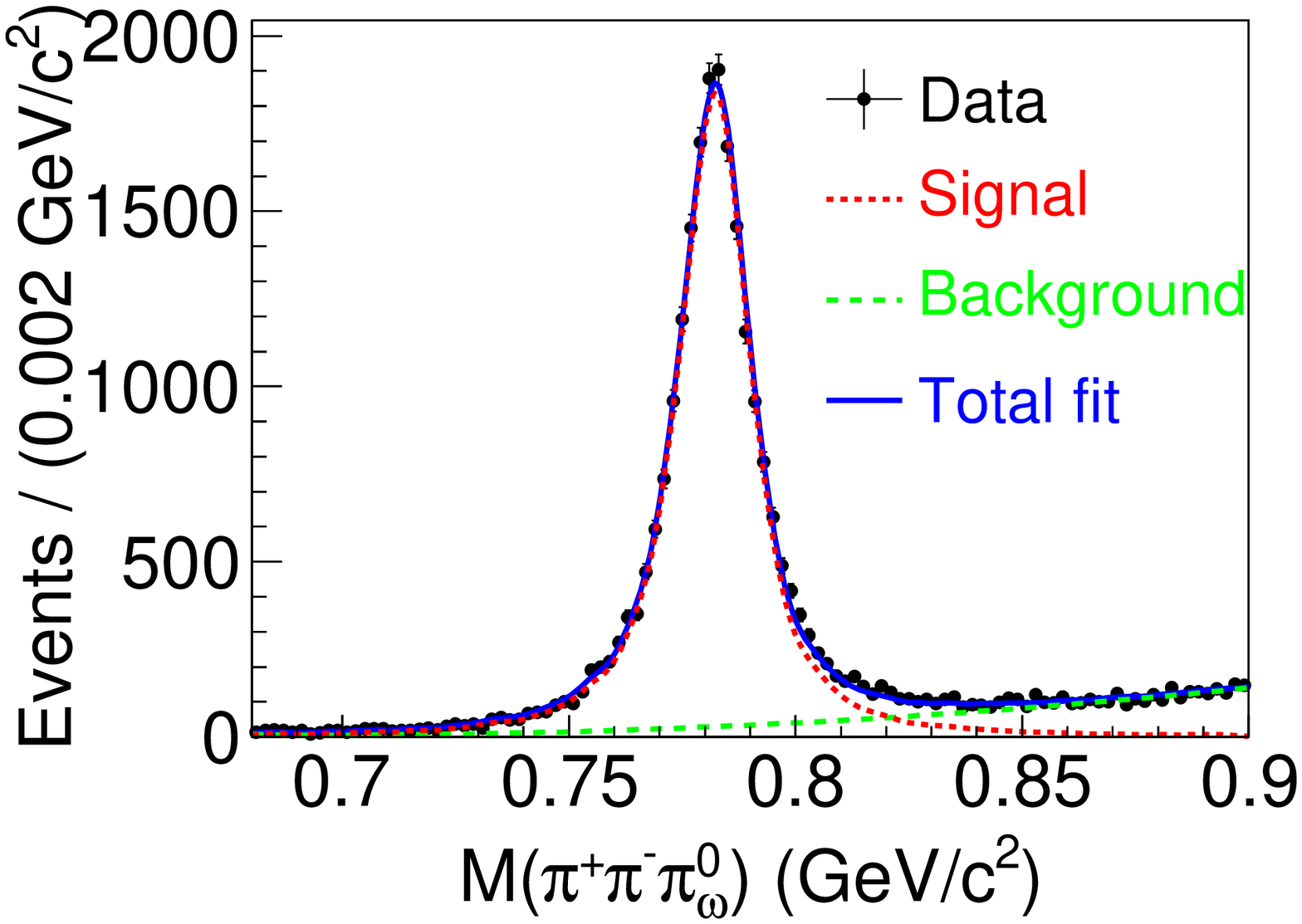}
\put(28,60){(c)}
\end{overpic}
\begin{overpic}[width=0.46\textwidth]{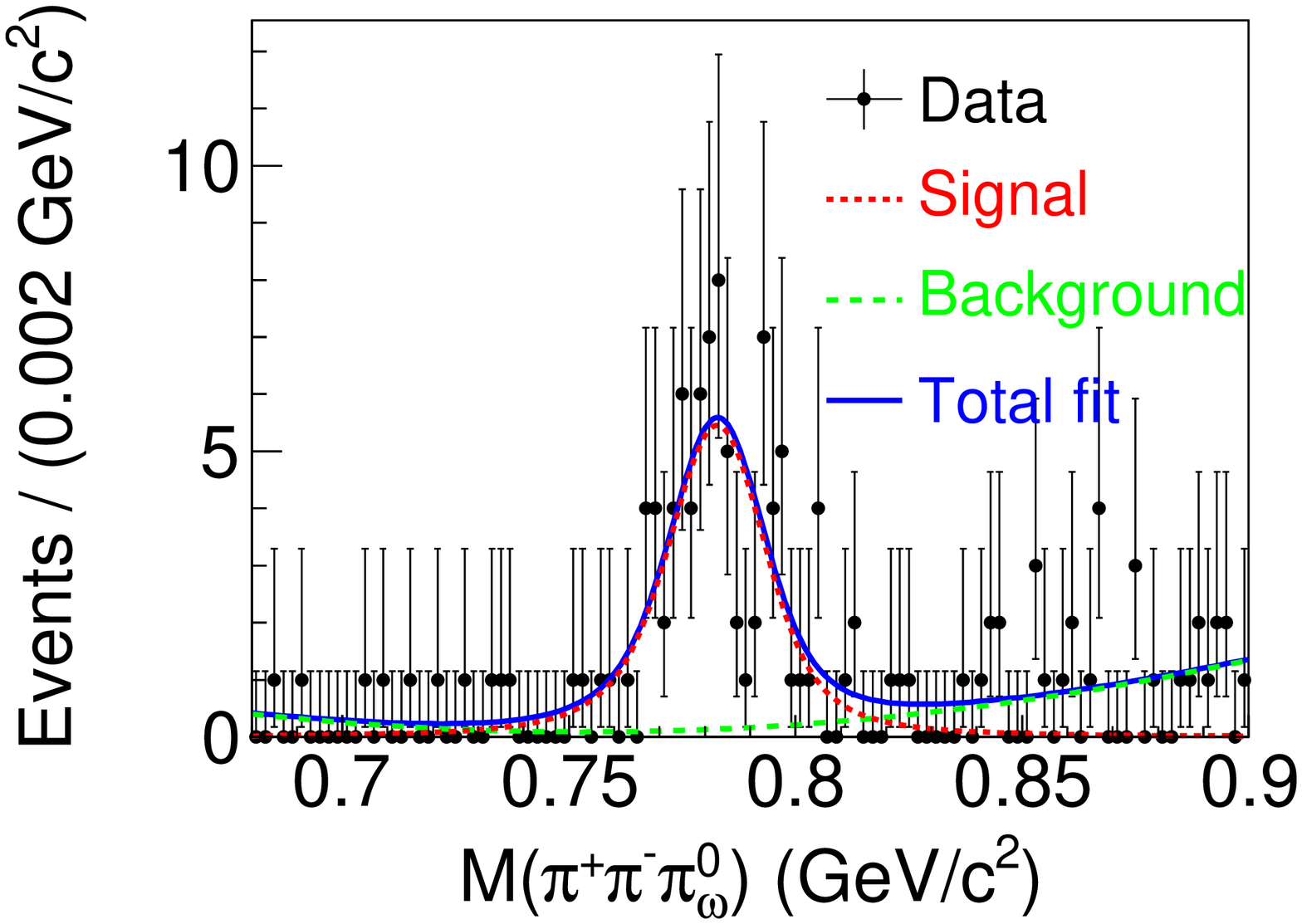}
\put(28,60){(d)}
\end{overpic}
\end{center}
\vspace{-0.5cm}
\caption{(Color online) Invariant mass distributions at $\sqrt{s}=2.125$ GeV. (a) Distribution of the $\pi^+\pi^-\pi^0_\omega$ invariant mass versus the two-photon invariant mass corresponding to the $\pi^0_{\mathrm{bach}}\to\gamma\gamma$ decay. 
The red box indicates the signal region.
(b) Distribution of the two-photon invariant mass $M(\gamma\gamma)$ corresponding to the $\pi^0_{\mathrm{bach}}\to\gamma\gamma$ decay, 
where the (black) dots with error bars are data, the (red) solid histogram and the (red) dashed histogram is the signal MC before and after $\pi^0_{\mathrm{bach}}$ matching with the MC truth information. The red and blue vertical lines indicate the signal and sideband regions, respectively.
(c) and (d) represent the $M(\pi^{+}\pi^{-}\pi^{0}_\omega)$ distribution corresponding to $\pi^{0}_{\mathrm{bach}}$ signal and sideband regions, respectively. The (black) dots with error bars are data, the (blue) solid curves are the total fit results, the (green) dashed curves indicate the background contributions described by a second order Chebychev polynomial and the (red) dotted curves show the $\omega$ signal shapes described by the MC lineshape convolved with a Gaussian function.}\label{OPomegaFit}
\end{figure}

Using the above selection criteria, the distribution of the invariant mass of $\pi^{+}\pi^{-}\pi^0_\omega$ versus the two-photon invariant mass for $\pi^0_\mathrm{bach}$ candidates is depicted in Fig.~\ref{OPomegaFit}(a). The $\omega$ signal is clearly evident.

\end{multicols}

\begin{multicols}{2}

A method similar to that described in Sec.~\ref{selomegaeta} is used
to study possible background contributions.  According to the study,
the dominant background stems from the four body process $e^+e^- \to
\pi^+\pi^-\pi^0\pi^0$, which has the same final state particles as the
signal channel.  In a similar way as in the $\ee\to\omega\eta$ case,
possible peaking background contributions are inferred from the
$\pi^0_\mathrm{bach}$ sideband regions defined as $0.055 <
|M(\gamma_1\gamma_2)-m_{\pi^0}| < 0.095 \gevcc$ (as illustrated in
Fig.~\ref{OPomegaFit}(b)).  Note that due to mis-combination of
photons, a large fraction of the $\pi^0$ sideband is composed of
signal reactions.  Still, while a peaking sideband contribution is
found, its fraction is negligible (and would still have to be scaled
down in a similar procedure as described for the $\omega\eta$ process)
compared to the signal region as shown in Figs.~\ref{OPomegaFit} (c)
and (d). %In addition, this small peaking background would have to be
%scaled down.

% Nils: this is obviously wrong, because it is stated above that the peaking background from the sideband is neglected...
%The signal yield is determined using the $M(\pi^+\pi^-\pi^0_\omega)$
%mass spectra (as shown in Fig.~\ref{OPomegaFit}(c)) with the method
%described in Sec.~\ref{selomegaeta}. The fit yields $N_{\rm sig} =
%22627\pm180$ events.
The signal yield is determined using the $M(\pi^+\pi^-\pi^0_\omega)$
mass spectra (as shown in Fig.~\ref{OPomegaFit}(c)) with a similar method as
described in Sec.~\ref{selomegaeta}, with the difference being that peaking backgrounds are neglected, so that the fit reduces to a one-dimensional unbinned likelihood fit.
The fit yields $N_{\rm sig} = 22627\pm180$ events.

The Born cross section of the $e^+e^- \to \omega\pi^0$ process is
calculated using Eq.~(\ref{CXOE}), with the product of the branching
fractions determined by $\mathcal{B}=\mathcal{B}(\omega \to
\pi^+\pi^-\pi^0)\cdot\mathcal{B}^2(\pi^0 \to \gamma\gamma)=87.1\%$.
The values used in the calculation of the Born cross section of the
$e^+e^- \to \omega \pi^0$ process are listed in
Table~\ref{crossOmegaPi0}, together with the results at all
c.m. energies.  The results are consistent with most of the previous
measurements~\cite{SND2000,SND2003,SND2016, ND, CMD1,CMD2, DM2} but
with improved precision, however, there exists a small difference with
the BaBar measurement \cite{BABAR} at center-of-mass energies around
$2.1$ GeV.  A comparison is shown in Fig.~\ref{fig:comparison}(b).

%\subsection{Systematic uncertainties}
Concerning the systematic uncertainties, the contribution stemming
from the luminosity determination is common for the $\ee\to\omega\eta$
and $\ee\to\omega\pi^0$ reactions. Furthermore, for the
uncertainties relating to the detection efficiencies, the radiative
corrections, the fitting procedure and the branching fractions taken
from the literature, the same method is applied as previously stated
in Sec.~\ref{selomegaeta}.  In addition, the uncertainty arising from
the $\pi^{0}$ selection is obtained by varying the mass window
requirements for both $\pi^{0}_\omega$ and $\pi^{0}_\mathrm{bach}$ and
examining the changes in the resulting cross sections.  The total
systematic uncertainty of the determination of the Born cross section
is determined to be 6.7\% for $e^+e^- \to \omega\pi^{0}$ at
$\sqrt{s}=2.125$ GeV. The uncertainties at the other c.m. energies are
determined accordingly and are summarized in
Table~\ref{UncertaintyOP}.

%\end{multicols}

%\begin{minipage}[]{0.49\textwidth}

\begin{table}[H] %[htbp]
\setlength{\tabcolsep}{3pt}
\footnotesize
  \centering
  \caption{ The Born cross sections of the $e^{+}e^{-} \to \omega \pi^0$ process. 
The symbols are the same as those in Eq.~(\ref{CXOE}).
In the column of the Born cross section $\sigma$, the first uncertainty is statistical and the second one is systematic.} 
% \begin{tabular}{p{1.1cm}<{\centering} p{1.6cm}<{\centering} p{0.9cm}<{\centering} p{1cm}<{\centering}  p{1.9cm}<{\centering}}
\begin{tabular}{ccccc}
  \hline
  \hline
$\sqrt{s}$ (GeV) &$N_{\rm sig}$  &$\mathcal{L}$ (${\rm pb}^{-1}$)    &$\varepsilon\cdot(1+\delta)$ &$\sigma$ (pb)\\
\hline
2.0000  &1677   $\pm$50       &10.1   &0.202    & 946  $\pm$28   $\pm$ 70   \\
2.0500  &652    $\pm$31       &3.34   &0.205    &1086  $\pm$52   $\pm$ 73   \\
2.1000  &2614   $\pm$62       &12.2   &0.209    &1181  $\pm$28   $\pm$ 80   \\
2.1250  &22627  $\pm$180      &108    &0.211    &1136  $\pm$ 9   $\pm$ 76   \\
2.1500  &539    $\pm$28       &2.84   &0.213    &1021  $\pm$52   $\pm$ 55   \\
2.1750  &1840   $\pm$51       &10.6   &0.217    & 914  $\pm$26   $\pm$ 59   \\
2.2000  &2064   $\pm$54       &13.7   &0.218    & 791  $\pm$21   $\pm$ 54   \\
2.2324  &1508   $\pm$46       &11.9   &0.222    & 659  $\pm$20   $\pm$ 43   \\
2.3094  &1846   $\pm$51       &21.1   &0.223    & 452  $\pm$13   $\pm$ 30   \\
2.3864  &1601   $\pm$48       &22.5   &0.222    & 366  $\pm$11   $\pm$ 26   \\
2.3960  &4553   $\pm$80       &66.9   &0.222    & 352  $\pm$ 6   $\pm$ 19   \\
2.5000  &53.8   $\pm$8.2      &1.10   &0.228    & 247  $\pm$38   $\pm$ 18   \\
2.6444  &1335   $\pm$42       &33.7   &0.234    & 195  $\pm$ 6   $\pm$ 11   \\
2.6464  &1274   $\pm$41       &34.1   &0.233    & 184  $\pm$ 6   $\pm$ 12   \\
2.7000  &34.9   $\pm$6.5      &1.03   &0.238    & 163  $\pm$30   $\pm$ 10   \\
2.8000  &21.2   $\pm$6.3      &1.01   &0.239    & 101  $\pm$30   $\pm$ 7.0   \\
2.9000  &2096   $\pm$54       &105    &0.243    &  93.8$\pm$ 2.4 $\pm$ 5.3  \\
2.9500  &302    $\pm$20       &15.9   &0.244    &  89.0$\pm$ 5.8 $\pm$ 5.2  \\
2.9810  &254    $\pm$19       &16.0   &0.246    &  74.0$\pm$ 5.5 $\pm$ 4.1  \\
3.0000  &256    $\pm$18       &15.9   &0.244    &  76.1$\pm$ 5.3 $\pm$ 4.1  \\
3.0200  &268    $\pm$18       &17.3   &0.242    &  73.3$\pm$ 5.0 $\pm$ 4.3  \\
3.0800  &1513   $\pm$40       &126  &0.223    &  61.8$\pm$ 1.7 $\pm$ 4.1  \\
\hline
\hline
    \end{tabular}
    \label{crossOmegaPi0}
\end{table}

%\end{minipage}

\end{multicols}

\begin{table}[H]
\setlength{\tabcolsep}{9pt}
%\begin{sidewaystable}
  \centering
  \scriptsize
  \caption{  Summary of relative systematic uncertainties (in \%) associated with the luminosity ($\mathcal{L}$),  the tracking efficiency (Track), the photon detection efficiency (Photon), PID, branching fraction (Br), 4C kinematic fit (4C), background shape (Bkg), signal shape (Sig), fit range (Range), $\pi^0$ mass windows (m($\pi^0$) and m($\pi^0_{\omega}$)), the initial state radiation and the vacuum polarization correction factor ($1+\delta$) in the measurement of the Born cross section of the  $e^{+}e^{-} \to \omega \pi^{0}$ process.}
  \label{UncertaintyOP}
  \begin{tabular}{ccccc ccccc cccc}
  \hline
  \hline
   Ecm  &   $\mathcal{L}$   &    Track & Photon  &  PID   &  Br    &  4C  &    Bkg   &     Sig  &    Range   &    $m(\pi^0)$   & $m(\pi^0_\omega)$  &   $(1+\delta)$  &   Total \\
%Energies &L &Track &Photon &PID &Br &$\chi^{2}$ &4C &$\cos(\theta)$  &Bkg.fit &Sig.fit  &Fit range    & m($\eta$)    &m($\pi^{0}$)    &Peaking &ISR  &Total    \\
\hline 
 2.0000  & 1.0  & 2.0  & 4.0  & 2.0  & 0.7  & 0.2  & 0.4  & 1.4  & 5.1  & 0.1  & 0.3  & 0.7  & 7.4  \\ 
 2.0500  & 1.0  & 2.0  & 4.0  & 2.0  & 0.7  & 0.3  & 0.4  & 0.3  & 4.4  & 0.4  & 0.4  & 0.6  & 6.8  \\ 
 2.1000  & 1.0  & 2.0  & 4.0  & 2.0  & 0.7  & 0.3  & 0.4  & 1.6  & 4.2  & 0.1  & 0.1  & 0.5  & 6.8  \\ 
 2.1250  & 1.0  & 2.0  & 4.0  & 2.0  & 0.7  & 0.3  & 0.2  & 1.1  & 4.2  & 0.1  & 0.1  & 0.5  & 6.7  \\ 
 2.1500  & 1.0  & 2.0  & 4.0  & 2.0  & 0.7  & 0.4  & 0.9  & 1.5  & 0.7  & 0.4  & 0.2  & 0.5  & 5.4  \\ 
 2.1750  & 1.0  & 2.0  & 4.0  & 2.0  & 0.7  & 0.4  & 0.0  & 0.4  & 4.1  & 0.0  & 0.2  & 0.5  & 6.6  \\ 
 2.2000  & 1.0  & 2.0  & 4.0  & 2.0  & 0.7  & 0.4  & 0.1  & 0.8  & 4.4  & 0.1  & 0.2  & 0.5  & 6.8  \\ 
 2.2324  & 1.0  & 2.0  & 4.0  & 2.0  & 0.7  & 0.4  & 0.1  & 0.8  & 3.9  & 0.2  & 0.3  & 0.5  & 6.5  \\ 
 2.3094  & 1.0  & 2.0  & 4.0  & 2.0  & 0.7  & 0.3  & 0.5  & 0.4  & 4.1  & 0.2  & 0.4  & 0.5  & 6.6  \\ 
 2.3864  & 1.0  & 2.0  & 4.0  & 2.0  & 0.7  & 0.3  & 1.4  & 0.9  & 4.7  & 0.3  & 0.4  & 0.5  & 7.1  \\ 
 2.3960  & 1.0  & 2.0  & 4.0  & 2.0  & 0.7  & 0.3  & 0.4  & 1.4  & 1.5  & 0.1  & 0.4  & 0.5  & 5.5  \\ 
 2.5000  & 1.0  & 2.0  & 4.0  & 2.0  & 0.7  & 0.4  & 0.4  & 2.4  & 4.2  & 1.3  & 1.0  & 0.5  & 7.2  \\ 
 2.6444  & 1.0  & 2.0  & 4.0  & 2.0  & 0.7  & 0.3  & 0.6  & 0.6  & 2.6  & 0.3  & 0.2  & 0.5  & 5.8  \\ 
 2.6464  & 1.0  & 2.0  & 4.0  & 2.0  & 0.7  & 0.1  & 0.9  & 1.4  & 3.4  & 0.5  & 0.3  & 0.5  & 6.4  \\ 
 2.7000  & 1.0  & 2.0  & 4.0  & 2.0  & 0.7  & 0.6  & 2.7  & 1.5  & 2.9  & 1.2  & 1.4  & 0.5  & 6.9  \\ 
 2.8000  & 1.0  & 2.0  & 4.0  & 2.0  & 0.7  & 0.1  & 0.2  & 1.6  & 2.1  & 0.7  & 0.9  & 0.5  & 5.8  \\ 
 2.9000  & 1.0  & 2.0  & 4.0  & 2.0  & 0.7  & 0.6  & 1.1  & 1.7  & 1.2  & 0.2  & 0.6  & 0.5  & 5.7  \\ 
 2.9500  & 1.0  & 2.0  & 4.0  & 2.0  & 0.7  & 0.2  & 0.5  & 1.2  & 2.3  & 0.7  & 0.2  & 0.5  & 5.8  \\ 
 2.9810  & 1.0  & 2.0  & 4.0  & 2.0  & 0.7  & 0.3  & 0.4  & 1.9  & 0.5  & 0.1  & 0.8  & 0.5  & 5.5  \\ 
 3.0000  & 1.0  & 2.0  & 4.0  & 2.0  & 0.7  & 0.4  & 0.3  & 0.5  & 1.4  & 0.7  & 0.6  & 0.5  & 5.4  \\ 
 3.0200  & 1.0  & 2.0  & 4.0  & 2.0  & 0.7  & 0.2  & 0.6  & 1.2  & 2.1  & 1.2  & 0.8  & 0.5  & 5.8  \\ 
 3.0800  & 1.0  & 2.0  & 4.0  & 2.0  & 0.7  & 0.5  & 0.8  & 1.7  & 3.5  & 1.1  & 0.8  & 0.5  & 6.6  \\
  \hline
  \hline
    \end{tabular}
%\end{sidewaystable}
\end{table}
%%%%%%%%%%%%%%%%%%%%%%%

%\begin{multicols}{2}
\begin{multicols}{2}

\section{Line shape analysis}
\subsection{Analysis of the $e^+e^- \to \omega\eta$ process}
\label{fitomegaeta}
\begin{figure}[H] %[htbp]
\begin{center}
\begin{overpic}[width=0.47\textwidth]{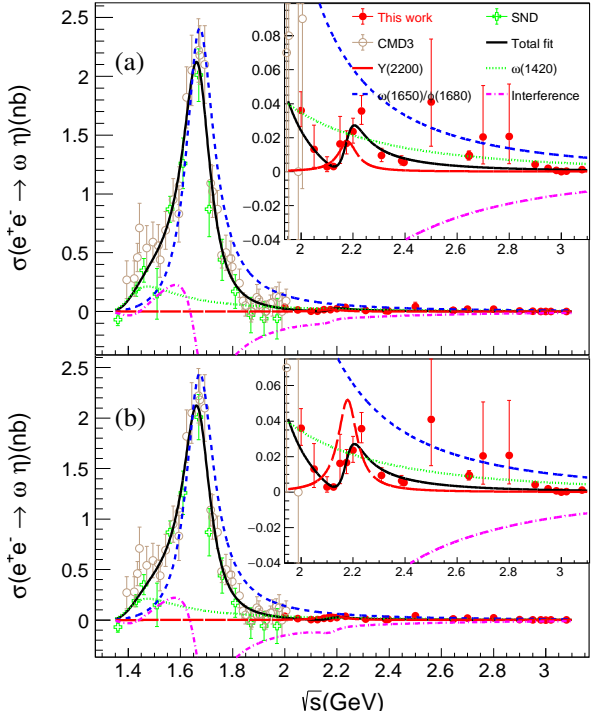}
\put(15,90){(a)}
\put(15,40){(b)}
\end{overpic}
\end{center}
\vspace{-0.5 cm}
\caption{(Color online) Fit to the dressed cross sections of $e^+e^- \to \omega \eta$. (a) Solution I. (b) Solution II. (Red) filled circles represent the data from this work, whereas (brown) open circles show the data from CMD3 and the (green) open crosses the data from SND. The (black) solid curves are the total fit results, the (red) long-dashed curves indicate the $Y(2180)$ resonance contribution, the (blue) short-dashed curves represent the $\omega(1650)$ or $\phi(1680)$ contribution, the (green) dotted curves display the $\omega(1420)$ contribution and (magenta) dotted-dashed curves show the interference contribution. In the upper right panel of both (a) and (b), a zoom into the region of the $Y(2180)$ resonance is shown.}\label{OmegaEtaFit}
\end{figure}

To study possible resonant structures in $e^+e^- \to \omega\eta$, 
%{\color{magenta} a minimum $\chi^2$ fit} 
 a maximum likelihood fit of the type used in Ref.~\cite{pipihc}
is performed to the dressed cross sections, which are the products of Born cross sections and VP factors. %with statistical uncertainties.
%{\color{red}
Previous results from the SND~\cite{SND2016omegaeta} and CMD3~\cite{CMD2017omegaeta} collaborations are also included to be able to describe the low-energy behavior of the cross section, while BaBar's result is not used due to their large uncertainties or non-observation without uncertainty.
%}
In the fit, a possible resonant amplitude is parameterized using a Breit-Wigner function with a mass-independent width. 
The flat contribution in the c.m. energy region between $2$ and $3$ GeV dominantly stems from tails of the $\omega(1420)$ and $\omega(1650)$ (or $\phi(1680)$) resonances. 
Following Ref.~\cite{SND2016omegaeta}, the dressed cross section is modeled as

\begin{equation}
%\vspace{-0.2cm}
%\begin{split}
\sigma(s)= \frac{12\pi}{s^{\frac{3}{2}}}\left| f_{1} - f_{2} + e^{i\varphi}f_{3} \right|^{2}P_{f}(s),\\
\label{eqOEcx}
%\end{split}
\end{equation}

\noindent
where $f_{R} = \sqrt{\frac{\Gamma^{ee}_{R}\cdot B^{\omega\eta}_{R}}{P_{f}(m_{R})}}\frac{m_{R}^{3/2}\sqrt{\Gamma_{R}}}{s - m^{2}_{R} + i\sqrt{s}\Gamma_{R}}$ (here $R=1,2,3$ is an index for the resonance) describes the resonant contributions from the $\omega(1420)$, $\omega(1650)$ (or $\phi(1680)$) and $Y(2180)$ (referring to the structure around $\sqrt{s}=2.2$ GeV) and
$\Gamma^{ee}_{R}\cdot B^{\omega\eta}_{R}$ is the product of the electronic width of the resonance $R$ and the branching fraction of the $R\to \omega\eta$ decay.
Furthermore, $m_{R}$ and $\Gamma_{R}$ are the mass and width of the resonance $R$, and
$\varphi$ is the relative phase angle of the $f_{3}$ contribution relative to the $f_1-f_2$ contribution.
The phase space factor $P_{f}(s)$ is given by $P_{f}(s) = q^{3}$, 
where $q$ is the $\omega$ momentum in the $e^+e^-$ c.m. frame
calculated for the mass value $m(\omega)=0.78265\gevcc$ given in Ref.~\cite{PDG}.
%\color{red}
The free fit parameters are taken as $\Gamma^{ee}\cdot B_{1}^{\omega\eta}$, $m_{2}$, $\Gamma_{2}$, $\Gamma^{ee}\cdot B_{2}^{\omega\eta}$, $m_{3}$, $\Gamma_{3}$,  $\Gamma^{ee}\cdot B_{3}^{\omega\eta}$ and ${\varphi}$. 
The $m_{1}$ and $\Gamma_{1}$ values are fixed to the values determined by the SND Collaboration~\cite{SND2016omegaeta}, 
since the significance of the $\omega(1420)$ resonance is not large enough at the given c.m. energies.
%In the fit, this measurement and previous results from the
%SND~\cite{SND2016omegaeta} and CMD3~\cite{CMD2017omegaeta} collaborations are included to be able to describe the low-energy behavior of the cross section. 
%{\color{magenta}
In the fit, 
uncertainties from previous experiments are considered uncorrelated, while 
the uncertainties derived in this work are split into the uncorrelated and the correlated contributions. The former contributions include those stemming from the choice of signal and background shape as well as fit range and the treatment of peaking backgrounds whereas the latter include the remaining systematic uncertainties. %}
Figure \ref{OmegaEtaFit} and Table~\ref{Table-TwoSol} show the results from our fit. Two solutions are found with the same fit quality of $\chi^2/\text{ndf} = 78/67$, where ndf is the number of degrees of freedom.
Solution I corresponds to constructive interference between the $f_3$ amplitude and the remaining $f_1-f_2$ contribution, while solution II corresponds to the case of destructive interference.
The two solutions share all parameters other than those given in Table~\ref{Table-TwoSol}.
%{\color{red} 
Among the other free parameters, the mass and width of $f_2$ are determined to be $1670\pm 4~\text{MeV}/c^2$ and $129\pm 7~\text{MeV}$, respectively, with $\Gamma^{ee}\cdot B_{f_2}^{\omega\eta}$ equal to $59\pm2~\text{eV}$.
%}

\begin{table}[H] %[htbp]
%\scriptsize
\footnotesize
%\small
\centering
\caption{Resonance parameters of the $Y(2180)$ as obtained in the fit to the $e^+e^- \to \omega \eta$ dressed cross section.}
  \begin{tabular}{p{2.8cm}cc}
  \hline
  \hline
  parameters                           %            &Value(\%)                     & \\
                                               & solution I  & solution II \\
  \hline
  $m_{Y(2180)} (\mevcc)$                       &\multicolumn{2}{c}{ $2179 \pm 21 $}    \\
  $\Gamma_{Y(2180)} (\mev)$                    &\multicolumn{2}{c}{ $89   \pm 28 $}    \\ 
  $\Gamma^{ee}\cdot B^{\omega\eta} (\ev)$              & $0.50 \pm 0.16$   & $1.50 \pm 0.44 $ \\
  $\varphi$                                    & $2.7 \pm 0.3$  & $1.9 \pm 0.2$ \\
   significance                       &\multicolumn{2}{c}{ 6.1$\sigma$}    \\
   \hline
  \hline
    \end{tabular}
    \label{Table-TwoSol}
\end{table}

\subsection{Analysis of the $e^+e^- \to \omega\pi^0$ process}
A fit is performed to the dressed cross sections of $e^+e^- \to
\omega\pi^0$ using a similar method as described in
Sec.~\ref{fitomegaeta}. 
%{\color{red}
Previous results from the SND collaboration~\cite{SND2003,SND2016} are included in order to provide the low-energy contributions that will only appear as tails in the energy region under study.
BaBar's result is not used since there is an obvious bias compared to the result in this work in the overlap region, and others are not used due to their large uncertainties.
%}
Here, the fit model is parameterized as a
coherent sum of four Breit-Wigner functions, %~\cite{SND2016}

\begin{equation}
\sigma(s)= \frac{12\pi}{s^{\frac{3}{2}}}\left| f_{1} + e^{i\varphi_{1}}f_{2} + e^{i\varphi_{2}}f_{3}  + e^{i\varphi_{3}}f_{4}  \right|^{2}P_{f}(s),\\
\end{equation}

\noindent
where $f_{R}$ (with $R = $ 1, 2, 3, 4) are the Breit-Wigner functions
for the $\rho(770)$, $\rho(1450)$, $\rho(1700)$ and $Y(2040)$
(referring to the structure around $\sqrt{s}=2.040$ GeV) resonances,
which take the the same form as described in Eq.~(\ref{eqOEcx}) except
for the $\rho(770)$.  Since the mass of the $\rho(770)$ resonance is
below the $\omega \pi^{0}$ threshold, we instead use $f_{\rho(770)} =
\frac{A}{s - m^{2}_{\rho(770)} + i\sqrt{s}\Gamma_{\rho(770)}(s)}$.
The formula for the energy-dependent width $\Gamma_{\rho(770)}(s)$ is
given in Ref.~\cite{SND2000}.  The free fit parameters are taken as
$A$, $\Gamma_{\rho(1450)}$, $\Gamma_{\rho(1700)}$, $m_{Y(2040)}$,
$\Gamma_{Y(2040)}$, $\Gamma^{ee}_{R}\cdot B^{\omega\eta}_{R}$ and
$\varphi_{R}$.

The masses of the $\rho(1450)$ and $\rho(1700)$ resonances are fixed to the average values as given by the PDG~\cite{PDG}.
In the fit,
a possible effect of omitting other data available in the literature on the results obtained in this work is studied and will be discussed in Sec.~\ref{sec:sys}.
Correlated and uncorrelated uncertainties of the present work are incorporated in the same way as described in Sec.~\ref{fitomegaeta}, while the uncertainties of the previous experiments are considered uncorrelated.

\begin{figure}[H]
\begin{center}
\begin{overpic}[width=0.49\textwidth]{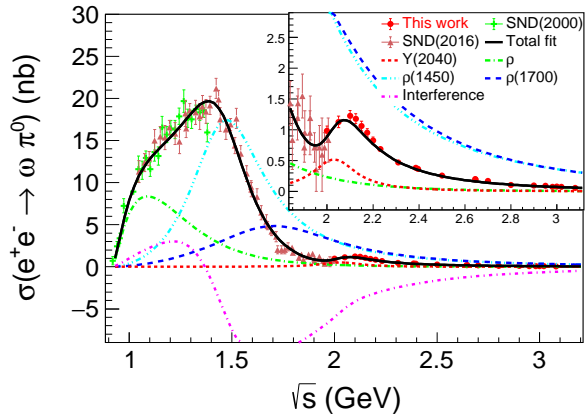}
%\put(20,60){(b)}
\end{overpic}
\end{center}
\vspace{-0.7 cm}
\caption{(Color online) Fit to the dressed cross sections of the $e^+e^- \to \omega \pi^0$ process. (Red) filled circles correspond to the data obtained in this work, while (brown) filled triangles and (green) filled crosses are the data from SND. The (black) solid curve is the total fit result, the (red) dashed curve is the $Y(2040)$ contribution, the (blue) long-dashed curve is the contribution from the $\rho(1700)$, the (light blue) dotted-dotted-dashed curve stems from the $\rho(1450)$, the (green) dotted-dashed curve corresponds to the $\rho(770)$ and the (magenta) dotted curve is the interference contribution.}\label{OmegaPi0Fit}
\end{figure}

The fit shown in Fig.~\ref{OmegaPi0Fit} finds a resonance with a mass of $(2034 \pm 13) \mevcc$, width of $(234 \pm 30)\mev$ and 
 $\Gamma^{ee}\cdot B^{\omega\pi^0}$ of $(34\pm11)~\text{eV}$
with a fit quality  of $\chi^{2}/\text{ndf} = 128/90$.
The significance of the $Y(2040)$ contribution is found to be larger than 10$\sigma$.

%width=6.0cm,height=4.5cm,angle=0

\subsection{Systematic uncertainties} \label{sec:sys}
The systematic uncertainties of the resonant parameters in the fit to the Born cross sections of $e^+e^- \to \omega\eta$ include contributions from the determination of the c.m. energy and the energy spread, fixed parameters in the fit, and the data from other experiments that is included in the fit. 
The uncertainty of the c.m. energy from BEPCII is small and found to be negligible comparing to the statistic uncertainty in the determination of the resonance parameters.
The effect resulting from fixing the parameters of the $\omega(1420)$ resonance is studied by varying the mass and width within the uncertainties quoted in the PDG~\cite{PDG} and yields an uncertainty of $\Delta m=3$ $\mevcc$, $\Delta \Gamma =$ 5 $\mev$ and $\Delta(\Gamma^{ee}\cdot B^{\omega\eta})$ equal to $0.03~\mbox{\rm eV}$ for solution I and $0.16~\mbox{\rm eV}$ for solution II.

We distinguish between two different types of systematic uncertainties, those that are uncorrelated between the different center-of-mass energies and those that are correlated.
While the uncorrelated uncertainties are included in the fit to the cross section, the correlated uncertainties that are common for all center-of-mass energies ($\sim 6\%$) only affect the $\Gamma^{ee}\cdot B^{\omega\eta}$ measurement and we find a resulting systematic uncertainty of $0.03~\mbox{\rm eV}$ for solution I and $0.09~\mbox{\rm eV}$ for solution II.
Assuming all sources of systematic uncertainties are uncorrelated and thus adding them in quadrature, the total systematic uncertainty is $3\mevcc$ for the mass, $5\mev$ for the width, $0.04~\mbox{\rm eV}$ (solution I) or $0.18~\mbox{\rm eV}$ (solution II) for $\Gamma^{ee}\cdot B^{\omega\eta}$ of the $Y(2180)$.

For the systematic uncertainties of the resonant parameters of the $Y(2040)$ in $e^+e^- \to \omega\pi^{0}$, 
the contribution introduced by taking the data points of other experiments into account in the fit is significant.
It is investigated by including all available measurements~\cite{SND2000,SND2003,SND2016,ND,CMD1,CMD2,BABAR,DM2} and comparing with the nominal fit result above.  
Other uncertainties are considered in the same way as stated before for the $Y(2180)\to\omega\eta$ case.
All sources of systematic uncertainties are added in quadrature, obtaining the total systematic uncertainty of $9\mevcc$ for the mass, $25\mev$ for the width and $16~\text{eV}$ for $\Gamma^{ee}\cdot B^{\omega\pi^0}$ of the observed $Y(2040)$.

\section{Summary and discussion}
The Born cross sections of the $e^+e^- \to \omega\eta$ and $e^+e^- \to \omega\pi^{0}$ processes have been measured at $\sqrt{s}$ from 2.000 to 3.080$\gev$. 
%They are consistent with previous measurements but with improved precision.
%{\color{red}
They are consistent with most of previous measurements in the overlap region, but deviate with BaBar's results, especially in the $\omega\pi^{0}$ process.
%}
Two resonant structures are observed in the measured line shapes.
One resonant structure is observed with a significance of 6.1$\sigma$ in the cross section of the $e^+e^- \to \omega\eta$ process,
with mass $m = (2179 \pm21 \pm3)\mevcc$, width $\Gamma = (89 \pm 28 \pm5)\mev$, 
and $\Gamma^{ee}\cdot B^{\omega\eta} = (0.50 \pm 0.16 \pm 0.04)~{\rm eV}$ or $(1.50 \pm 0.44 \pm 0.18)~{\rm eV}$, depending on the choice between two ambiguous fit solutions. 
The observed structure agrees well with the properties of the $\phi(2170)$ resonance, which indicates the first observation of the decay $\phi(2170)\to\omega\eta$.

Another structure is observed in the $\omega\pi^0$ cross section with a significance of more than 10$\sigma$ and
with a mass of $m = (2034 \pm 14 \pm 9)\mevcc$, width of $\Gamma = (234 \pm 30 \pm 25)\mev$ and $\Gamma^{ee}\cdot B^{\omega\pi^0}$ of $(34\pm11\pm16)~\text{eV}$. 
This structure could either be the $\rho(2000)$ or the $\rho(2150)$ state.
However, the mass and width of the observed resonance is closer to the $\rho(2000)$ resonance, which is suggested to be the $2^{3}D_{1}$ state~\cite{Rhoresonance}.
%The nature of this $Y(2040)$ resonance can however not be clarified here -- further efforts and more detailed investigations are needed.

%%%%%%%%%%%%%%%%%%%%%%%%%%%%%%%%%%%%%%%%%%%%%%%%%%%%%%%%%%%%%%%%%%%%%%%%%%%%%%%%%%%%%%%%%%%%%%%%%%%%%%%%%%
\section*{Acknowledgements}

The BESIII collaboration thanks the staff of BEPCII, the IHEP computing center and the supercomputing center of USTC for their strong support. 
This work is supported in part by National Key Basic Research Program of China under Contract No. 2015CB856700; 
National Natural Science Foundation of China (NSFC) under Contracts Nos. 11335008, 11375170, 11475164, 11475169, 11625523, 11605196, 11605198, 11635010, 11705192, 11735014, 11822506, 11835012, 11935015, 11935016, 11935018, 11950410506, 11961141012, 12035013; 
%11335008，11375170, 11475164， 11475169，11625523, 11605196，11605198, 11705192, 
the Chinese Academy of Sciences (CAS) Large-Scale Scientific Facility Program; 
Joint Large-Scale Scientific Facility Funds of the NSFC and CAS under Contracts Nos. U1532102, U1732263, U1832103, U1832207, U2032111; 
CAS Key Research Program of Frontier Sciences under Contracts Nos. QYZDJ-SSW-SLH003, QYZDJ-SSW-SLH040; 
100 Talents Program of CAS; 
INPAC and Shanghai Key Laboratory for Particle Physics and Cosmology; 
ERC under Contract No. 758462; 
German Research Foundation DFG under Contracts Nos. Collaborative Research Center CRC 1044, FOR 2359; 
Istituto Nazionale di Fisica Nucleare, Italy; 
Ministry of Development of Turkey under Contract No. DPT2006K-120470; 
National Science and Technology fund; 
STFC (United Kingdom); 
The Knut and Alice Wallenberg Foundation (Sweden) under Contract No. 2016.0157; 
The Royal Society, UK under Contracts Nos. DH140054, DH160214; 
The Swedish Research Council; 
U. S. Department of Energy under Contracts Nos. DE-FG02-05ER41374, DE-SC-0012069.

%The BESIII collaboration thanks the staff of BEPCII and the IHEP computing center for their strong support. This work is supported in part by National Key Basic Research Program of China under Contract No. 2015CB856700; National Natural Science Foundation of China (NSFC) under Contracts Nos. 11625523, 11635010, 11735014, 11822506, 11835012, 11935015, 11935016, 11935018, 11961141012; the Chinese Academy of Sciences (CAS) Large-Scale Scientific Facility Program; Joint Large-Scale Scientific Facility Funds of the NSFC and CAS under Contracts Nos. U1732263, U1832207; CAS Key Research Program of Frontier Sciences under Contracts Nos. QYZDJ-SSW-SLH003, QYZDJ-SSW-SLH040; 100 Talents Program of CAS; INPAC and Shanghai Key Laboratory for Particle Physics and Cosmology; ERC under Contract No. 758462; German Research Foundation DFG under Contracts Nos. Collaborative Research Center CRC 1044, FOR 2359; Istituto Nazionale di Fisica Nucleare, Italy; Ministry of Development of Turkey under Contract No. DPT2006K-120470; National Science and Technology fund; STFC (United Kingdom); The Knut and Alice Wallenberg Foundation (Sweden) under Contract No. 2016.0157; The Royal Society, UK under Contracts Nos. DH140054, DH160214; The Swedish Research Council; U. S. Department of Energy under Contracts Nos. DE-FG02-05ER41374, DE-SC-0012069

\end{multicols}


\section*{References}
\begin{thebibliography}{99}
%\bibitem{CQM}E. Santopinto {\it et al.}, AIP Conference Proceedings 1560, 451 (2013).

\bibitem{PDG} P.A. Zyla {\it et al.} (Particle Data Group),
  Prog. Theor. Exp. Phys. 2020, 083C01 (2020).

%\bibitem{PDG} C. Patrignani {\it et al.} (Particle Data Group), Chin.\ Phys.\ C.\ {\bf 40}, 100001 (2016).
%%%%%%%%\phi 2170 %%%%%%%%%%%%%%%%%%

\bibitem{ssg_phi2170} G. J. Ding and M. L. Yan, Phys.\ Lett.\ B {\bf 650}, 390 (2007).
\bibitem{ssg2_phi2170} J. Ho, R. Berg, and T. G. Steele, Phys.\ Rev.\ D {\bf 100}, 034012 (2019).
\bibitem{ssbar_phi2170} G. J. Ding and M. L. Yan, Phys.\ Lett.\ B {\bf 657}, 49 (2007).
\bibitem{ssbar2_phi2170} C. Q. Pang, Phys.\ Rev.\ D {\bf 99}, 074015 (2019).
\bibitem{ssbar3_phi2170} C. G. Zhao {\it et al.}, Phys.\ Rev.\ D {\bf 99}, 114014 (2020).
\bibitem{ssbar4_phi2170} Q. Li {\it et al.}, arXiv: 2004.05786.

\bibitem{sssbarsbar1_phi2170} Z. G. Wang, Nucl.\ Phys.\ A {\bf 791}, 106 (2007).
\bibitem{sssbarsbar2_phi2170} C. R. Deng, J. L. Ping, and T. Goldman, Phys.\ Rev.\ D\ {\bf 82}, 074001 (2010).
\bibitem{sssbarsbar3_phi2170} S. S. Agaev, K. Azizi, and H. Sundu, Phys.\ Rev.\ D\ {\bf 101}, 074012 (2020).
\bibitem{sssbarsbar4_phi2170} H. W. Ke and X. Q. Li, Phys.\ Rev.\ D\ {\bf 99}, 036014 (2019).
\bibitem{sssbarsbar5_phi2170} R. R. Dong {\it et al.}, Eur.\ Phys.\ J.\ C\ {\bf 80}, 749 (2020).
\bibitem{sssbarsbar6_phi2170} F. X. Liu {\it et al.}, arXiv: 2008.01372.

\bibitem{Lambda1_phi2170} L. Zhao {\it et al.}, Phys.\ Rev.\ D\ {\bf 87}, 054034 (2013).
\bibitem{Lambda2_phi2170} E. Klempt and A. Zaitsev, Phys.\ Rep.\ {\bf 454}, 1 (2007).
\bibitem{Lambda3_phi2170} Y. L. Yang, D. Y. Chen, and Z. Lu, Phys.\ Rev.\ D\ {\bf 100}, 073007 (2019).
\bibitem{phiKK_phi2170} A. M. Torres {\it et al.}, Phys.\ Rev.\ D\ {\bf 78}, 074031 (2008).
\bibitem{phif02_phi2170} L. Alvarez-Ruso, J. A. Oller, and J. M. Alarc$\acute{o}$n, Phys.\ Rev.\ D\ {\bf 80}, 054011 (2009).

\bibitem{quarkonia} T. Barnes N. Black, and P. R. Page, Phys.\ Rev.\ D {\bf 68}, 054014 (2003). 
\bibitem{LLbarbound} Y. Dong {\it et al.}, Phys.\ Rev.\ D\ {\bf 96}, 074027 (2017).
\bibitem{phietatheory} S. S. Agaev, K. Azizi, and H. Sundu, Phys.\ Rev.\ D\ {\bf 101}, 074012 (2020).

\bibitem{BABAR_phi2170} B. Aubert {\it et al.} (BABAR Collaboration), Phys.\ Rev.\ D\ {\bf 74}, 091103(R) (2006).
\bibitem{BESII_phi2170} M. Ablikim {\it et al.} (BES Collaboration), Phys.\ Rev.\ Lett.\ {\bf 100}, 102003 (2008).
\bibitem{BESIII_phi2170} M. Ablikim {\it et al.} (BESIII Collaboration), Phys.\ Rev.\ D\ {\bf 91}, 052017 (2015).
\bibitem{Belle_phi2170} C. P. Shen {\it et al.} (Belle Collaboration), Phys.\ Rev.\ D\ {\bf 80}, 031101(R) (2009).
\bibitem{BABAR2_phi2170} J. P. Lees {\it et al.} (BABAR Collaboration), Phys.\ Rev.\ D\ {\bf 86}, 012008 (2012).

\bibitem{BESKKpipi} M.~Ablikim {\it et al.} (BESIII Collaboration), 	Phys.\ Rev.\ Lett.\ {\bf 124}, 112001 (2020).
\bibitem{BESKK} M.~Ablikim {\it et al.} (BESIII Collaboration), Phys.\ Rev.\ D {\bf 99}, 032001 (2019).
\bibitem{BESKKTheory} D. Y. Chen, J. Liu, and J. He,  Phys.\ Rev.\ D\ {\bf 101}, 074045 (2020).
\bibitem{BESphietap} M.~Ablikim {\it et al.} (BESIII Collaboration), Phys.\ Rev.\ D\ {\bf 102}, 012008 (2020).
\bibitem{BESphiKK} M.~Ablikim {\it et al.} (BESIII Collaboration), Phys.\ Rev.\ D {\bf 100}, 032009 (2019).

\bibitem{omegafamily} C. Q. Pang {\it et al.}, Phys.\ Rev.\ D {\bf 101}, 074022 (2020).

\bibitem{rho1} A. Hasan and D. V. Bugg, Phys.\ Lett.\ B\ {\bf 334}, 215 (1994).
\bibitem{rho2} D. V. Bugg, Phys.\ Rept.\ {\bf 397}, 257 (2004).
\bibitem{babar2020} J. P. Lees {\it et al.} (BABAR Collaboration), Phys.\ Rev.\ D {\bf 101}, 012011 (2012).
\bibitem{rho3} J. P. Lees {\it et al.} (BABAR Collaboration), Phys.\ Rev.\ D {\bf 86}, 032013 (2012).
\bibitem{rho4} B. Aubert {\it et al.} (BABAR Collaboration), Phys. Rev. D {\bf 76}, 092005 (2007).
\bibitem{rho5} M. E. Biagini, S. Dubnicka and E. Etim, Il Nuovo Cimento, Vol. {\bf 104} A, N. 3 (1991).
\bibitem{rho6} A. B. Clegg and A. Donnachie, Z. Phys. C - Particles and Fields {\bf 45}, 677 (1990).

%\bibitem{phieta_babar} B. Aubert, {\it et al.}, (BABAR Collaboration), Phys.\ Rev.\ D\ {\bf 77}, 092002 (2008).
\bibitem{Rhoresonance2} L. M. Wang, J. Z. Wang and X. Liu, Phys.\ Rev.\ D {\bf 102}, 034037 (2020).
\bibitem{Rhoresonance} L. P. He, X. Wang and X. Liu, Phys.\ Rev.\ D {\bf 88}, 034008 (2013).
\bibitem{SND2000} M.N. Achasov {\it et al.} (SND Collaboration), Phys.\ Lett.\ B {\bf 486}, 29 (2000).
\bibitem{SND2003} M.N. Achasov {\it et al.} (SND Collaboration), J.\ Exp.\ Theor.\ Phys.\ {\bf 96}, 789 (2003).
\bibitem{SND2016} M.N. Achasov {\it et al.} (SND Collaboration), Phys.\ Rev.\ D {\bf 94}, 112001 (2016).
\bibitem{ND} S.I. Dolinsky {\it et al.} (ND Collaboration), Phys.\ Lett.\ B {\bf 174}, 453 (1986).
\bibitem{CMD1} R.R. Akhmetshin {\it et al.} (CMD-2 Collaboration), Phys.\ Lett.\ B {\bf 466}, 392 (1999).
\bibitem{CMD2} R.R. Akhmetshin {\it et al.} (CMD-2 Collaboration), Phys.\ Lett.\ B {\bf 562}, 173 (2003).
\bibitem{DM2} D. Bisello, {\it et al.} (DM2 Collaboration), Orsay preprint LAL 90-35 (1990): contributed paper to the International Conference on High Energy Physics, Singapore, 1990.
\bibitem{BABAR} J. P. Lees {\it et al.} (BABAR Collaboration), Phys.\ Rev.\ D {\bf 96}, 092009 (2017). %arXiv:1709.01171.
\bibitem{Ablikim:2009aa} M.~Ablikim {\it et al.} (BESIII Collaboration),  Nucl.\ Instrum.\ Meth.\ A {\bf 614}, 345 (2010).
\bibitem{Yu:IPAC2016-TUYA01} C.~H.~Yu {\it et al.}, Proceedings of IPAC2016, Busan, Korea, 2016, doi:10.18429/JACoW-IPAC2016-TUYA01.
%\bibitem{etof} X.~Li {\it et al.}, Radiat. Detect. Technol. Methods {\bf 1}, 13 (2017); Y.~X.~Guo {\it et al.}, Radiat. Detect. Technol. Methods {\bf 1}, 15 (2017).
\bibitem{geant4} S.~Agostinelli {\it et al.} (GEANT4 Collaboration), Nucl.\ Instrum.\ Meth.\ A {\bf 506}, 250 (2003).
\bibitem{boost} Z. Y. Deng {\it et al.} Chin.\ Phys.\  C {\bf 30}, 371 (2006).
\bibitem{ConExc} R. G. Ping, Chin.\ Phys.\  C {\bf 38}, 083001 (2014).

%\bibitem{BESIII} M. Ablikim {\it et al.} (BESIII Collaboration), Phys.\ Rev.\ D.\ {\bf 91}, 112008 (2015).
%\bibitem{BESIIIdetector} M. Ablikim {\it et al.} (BESIII Collaboration), Nucl.\ Instrum.\ Methods.\ Phys.\ Res.\ Sect.\ A {\bf 614}, 345 (2010).
%\bibitem{KKMC} S. Jadach, B. F. L. Ward and Z. Was, Comput.\ Phys.\ Commun.\ {\bf 130}, 260 (2000); Phys.\ Rev.\ D {\bf 63}, 113009 (2001).
%\bibitem{BESEVTGEN} D. M. Asner {\it et al.} Int.\ J.\ Mod.\ Phys.\ A {\bf 24}, S1 (2009); R. G. Ping, Chin.\ Phys.\ C {\bf 32}, 599 (2008); D. J. Lange, Nucl.\ Instrum.\ Methods Phys.\ Res.\, Sect.\ A {\bf 462}, 152 (2001).
\bibitem{SND2016omegaeta} M.N. Achasov {\it et al.} (SND Collaboration), Phys.\ Rev.\ D {\bf 94}, 092002 (2016).
\bibitem{CMD2017omegaeta} R.R. Akhmetshin {\it et al.} (CMD-3 Collaboration), Phys.\ Lett.\ B {\bf 773}, 150 (2017).
\bibitem{BaBarOmegaEta} B. Aubert {\it et al.} (BABAR Collaboration), Phys.\ Rev.\ D {\bf 73}, 052003 (2006).
\bibitem{Luminosity} M. Ablikim {\it et al.} (BESIII Collaboration), Chin. Phys. C {\bf 41}, 063001 (2017).
\bibitem{Tracking} W. L. Yuan {\it et al.}, Chin. Phys. C {\bf 40}, 026201 (2016).
%\bibitem{BESIIIPre} http://hnbes3.ihep.ac.cn/HyperNews/get/paper271.html.
%\bibitem{ISRcorrection} M. Ablikim {\it et al.} (BESIII Collaboration), Phys.\ Rev.\ Lett.\ {\bf 110}, 252001 (2013).
\bibitem{4C} M. Ablikim {\it et al.} (BESIII Collaboration), Phys.\ Rev.\ D {\bf 87}, 012002 (2013).

%\bibitem{BABARphieta} B. Aubert {\it et al.} (BABAR Collaboration), Phys.\ Rev.\ D {\bf 77}, 092002 (2008).

%\bibitem{RhoRho3Xiaoyun} X. Chen and J. Ping, Phys.\ Rev.\ D {\bf 95}, 114014 (2017).
% Nils: I am removing this, it's not quoted anywhere...

\bibitem{pipihc} M. Ablikim {\it et al.} (BESIII Collaboration), Phys.\ Rev.\ Lett.\ {\bf 118}, 092002 (2017).

\end{thebibliography}
\end{document}